\documentclass[lettersize,journal]{IEEEtran}
\usepackage{cite}
\usepackage{amsmath,amsfonts}
\usepackage{graphicx}

\usepackage{textcomp}
\usepackage{xcolor}
\usepackage{CJKutf8}
\usepackage{tabularray}
\usepackage{subfigure}
\usepackage{caption}
\usepackage{multirow}
\usepackage{bm}
\usepackage{amsmath,amsfonts}
\usepackage{array}
\usepackage[caption=false,font=normalsize,labelfont=sf,textfont=sf]{subfig}
\usepackage{textcomp}
\usepackage{stfloats}
\usepackage{url}
\usepackage{verbatim}
\usepackage{graphicx}
\usepackage{cite}
\usepackage{amsthm}
\usepackage{cite}
\usepackage{amsmath,amssymb,amsfonts}
\usepackage{graphicx}
\usepackage{textcomp}
\usepackage{xcolor}
\usepackage{CJKutf8}
\usepackage{tabularray}
\usepackage{caption}
\usepackage{multirow}
\usepackage{bm}
\usepackage{amsmath,amsfonts}
\usepackage{array}
\usepackage[caption=false,font=normalsize,labelfont=sf,textfont=sf]{subfig}
\usepackage{textcomp}
\usepackage{stfloats}
\usepackage{url}
\usepackage{verbatim}
\usepackage{graphicx}
\def\BibTeX{{\rm B\kern-.05em{\sc i\kern-.025em b}\kern-.08em
    T\kern-.1667em\lower.7ex\hbox{E}\kern-.125emX}}
\usepackage{balance}
\usepackage{amsmath}
\allowdisplaybreaks[4]
\usepackage{cite}
\usepackage{amsmath,amsfonts}
\usepackage{graphicx}
\usepackage{textcomp}
\usepackage{xcolor}
\usepackage{CJKutf8}
\usepackage{tabularray}
\usepackage{caption}
\usepackage{multirow}
\usepackage{bm}
\usepackage{amsmath,amsfonts}
\usepackage{algorithm}
\usepackage{array}
\usepackage{subcaption}
\usepackage[caption=false,font=normalsize,labelfont=sf,textfont=sf]{subfig}
\usepackage{textcomp}
\usepackage{stfloats}
\usepackage{url} 
\usepackage{verbatim}
\usepackage{graphicx}
\usepackage{cite}
 \usepackage{subcaption}
\usepackage{amsmath,amsfonts}
\usepackage{array}
\usepackage[caption=false,font=normalsize,labelfont=sf,textfont=sf]{subfig}
\usepackage{textcomp}
\usepackage{stfloats}
\usepackage{url}
\usepackage{verbatim}
\usepackage{graphicx}
\def\BibTeX{{\rm B\kern-.05em{\sc i\kern-.025em b}\kern-.08em
    T\kern-.1667em\lower.7ex\hbox{E}\kern-.125emX}}
\usepackage{balance}
\usepackage{amsmath}
\usepackage{subcaption}
\allowdisplaybreaks[4]
\usepackage{multicol}
\usepackage{cuted}
\usepackage{multicol,amsmath}
\usepackage{makecell}
\usepackage{array}
\newcolumntype{C}{>{\centering\arraybackslash}X}
\usepackage{tabularx,booktabs}
\usepackage[flushleft]{threeparttable}

\usepackage{booktabs,ragged2e}
\usepackage{multirow}
\usepackage{subcaption}
\usepackage{bm}
\usepackage{amsthm}
\allowdisplaybreaks[4]

\begin{document}
\begin{CJK}{UTF8}{gbsn}
\title{Can Knowledge Improve Security? A Coding-Enhanced Jamming Approach for Semantic Communication}

\author{Weixuan Chen,~\IEEEmembership{Graduate Student Member, IEEE}, Qianqian Yang,~\IEEEmembership{Member, IEEE}, Shuo Shao,~\IEEEmembership{Member, IEEE}, Zhiguo Shi,~\IEEEmembership{Fellow, IEEE}, Jiming Chen,~\IEEEmembership{Fellow, IEEE}, and Xuemin (Sherman) Shen,~\IEEEmembership{Fellow, IEEE}
\thanks{W. Chen, Q. Yang$^{\dag}$, and Z. Shi are with the College of Information Science and Electronic Engineering, Zhejiang University, Hangzhou,
China. (e-mails: \{weixuanchen, qianqianyang20$^{\dag}$, shizg\}@zju.edu.cn). 

S. Shao is with the Department of System Science, University of Shanghai for Science and Technology, Shanghai, China. (e-mail: shuoshao@usst.edu.cn). 

J. Chen is with the College of Control Science and Engineering, Zhejiang University, Hangzhou, China. (e-mail: cjm@zju.edu.cn).

X. Shen is with the Department of Electrical and Computer Engineering, University of Waterloo, Waterloo, Canada. (e-mail: sshen@uwaterloo.ca).

This work is partly supported by the NSFC under grant No. 62293481, No. 62571487, No. 62201505, by the National Key R\&D Program of China under Grant 2024YFE0200802, and by the Zhejiang Provincial Natural Science Foundation of China under Grant No. LZ25F010001.  (Corresponding author: Qianqian Yang.)  

The implementation code is publicly available at: https://github.com/1weixuanchen/A-Coding-Enhanced-Jamming-Approach-for-Semcom.

}

}

\maketitle


\begin{abstract}
As semantic communication (SemCom) attracts growing attention as a novel communication paradigm, ensuring the security of transmitted semantic information over open wireless channels has become a critical issue. 
However, traditional encryption methods often introduce significant additional communication overhead to maintain reliability, 
and conventional learning-based secure SemCom methods typically rely on a channel capacity advantage for the legitimate receiver, 
which is challenging to guarantee in real-world scenarios.
In this paper, we propose a coding-enhanced jamming method that eliminates the need to transmit a secret key by utilizing shared knowledge, which may be part of the training set of the SemCom system, between the legitimate receiver and the transmitter.
Specifically, we leverage the shared private knowledge base to generate a set of private digital codebooks in advance using neural network (NN)-based encoders.
For each transmission, we encode the transmitted data into a digital sequence $\textbf{Y}_1$ and associate $\textbf{Y}_1$ with a sequence randomly picked from the private codebook, denoted as $\textbf{Y}_2$, through superposition coding.
Here, $\textbf{Y}_1$ serves as the outer code and
$\textbf{Y}_2$ as the inner code. 
By optimizing the power allocation between the inner and outer codes, the legitimate receiver can reconstruct the transmitted data using successive decoding based on the shared index of $\textbf{Y}_2$, while the eavesdropper's decoding performance is severely degraded, potentially to the point of random guessing.
Experimental results demonstrate that our method achieves security comparable to state-of-the-art approaches while significantly improving the reconstruction performance of the legitimate receiver by more than 1 dB across varying channel signal-to-noise ratios (SNRs) and compression ratios.


\end{abstract}

\begin{IEEEkeywords}
Semantic communications, encoded jammer, digital communications, superposition coding, wiretap channels.
\end{IEEEkeywords}



\section{Introduction}

Traditional communication systems prioritize bit-level accuracy in message transmission but often neglect semantic meaning.
This design results in a lack of relevance and flexibility, as the same transmission strategy is applied across different message contexts. 
Semantic communication (SemCom) \cite{zhang2024semanticsurvey} overcomes this limitation by focusing on the effective delivery of message semantics.
Specifically, SemCom approaches leverage the power of deep learning to selectively extract and transmit core semantic information while discarding redundant content.
At the core of this paradigm lies deep joint source-channel coding (DeepJSCC) \cite{bourtsoulatze2019deep}, a neural network (NN)-based approach that extracts task-relevant semantics from the source data and directly maps them to transmit symbol sequences. This method has been shown to outperform conventional communication techniques across various data transmission tasks, such as text \cite{han2022semantictext}, speech\cite{weng2023deep,han2022semanticpre}, image\cite{zhang2022semantic,chen2023deep,tang2025towards}, and video\cite{jiang2022wireless}.


Although SemCom offers superior performance, transmitting semantic information over open wireless channels introduces critical security vulnerabilities. 
For instance, in wiretap channels, an eavesdropper may attempt to recover the source data by intercepting the semantic information transmitted through the channel.
Recent efforts to develop secure SemCom systems over wiretap channels can be broadly classified into three categories based on the advantages of the legitimate user over the eavesdropper.

When the legitimate receiver has no advantage at all over the eavesdropper in terms of channel conditions, secure SemCom can still be achieved, but the communication rate may be very low, or the achievable security level may be limited. 
For instance, Erdemir \textit{et al.}'s variational autoencoder (VAE)-based JSCC framework \cite{erdemir2022privacy} extracts semantic information from source data and quantifies privacy leakage via mutual information (MI) between the sensitive information inferred by the eavesdropper and the eavesdropped semantic data.
They optimize the trade-off between reducing privacy leakage to potential eavesdroppers and maintaining high reconstruction performance for legitimate receivers. 
Similarly, in the context of adversarial training,
Marchioro \textit{et al.} \cite{marchioro2020adversarial} proposed a data-driven secure SemCom scheme using adversarial networks. 
They formulated a game between the legitimate receiver and the eavesdropper, aiming to maximize the reconstruction performance of the legitimate receiver while penalizing information leakage to the eavesdropper. 
Tung \textit{et al.} \cite{tung2023deep} proposed a secure SemCom scheme combining public-key cryptography \cite{lindner2011better} with DeepJSCC, effectively preventing eavesdropping and chosen-plaintext attacks while improving both security and image quality. 
However, these methods share a fundamental limitation: they treat security as a training penalty rather than enforcing explicit security control during transmission.

When there is no key-sharing channel but the legitimate receiver has better channel conditions than the eavesdropper, 
secure SemCom can still be achieved through physical layer security methods. 
For example, Chen \textit{et al.}\cite{chen2024nearly}
proposed a superposition coding-based approach that overlays a 4-ary quadrature amplitude modulation (4-QAM) constellation sequence encoding semantic information onto another randomly generated 4-QAM constellation sequence. 
This method enables explicit security control by adjusting the power allocation between the two sequences, effectively minimizing information leakage to eavesdroppers.
However, it assumes that the eavesdropping channel is significantly worse than the legitimate channel, which may not hold true in many practical scenarios.

When a secure key-sharing channel exists between the transmitter and the legitimate receiver, the transmitter can leverage this key to encrypt the transmitted signal.
For instance, in the encrypted SemCom system (ESCS) developed by Luo \textit{et al.} \cite{luo2023encrypted}, an NN-based encryptor with randomly generated keys is used to encrypt semantic information, applying adversarial training to prevent eavesdroppers from successfully recovering the source data.
While adversarial training enhances security in both encrypted and unencrypted modes, it only penalizes the eavesdropper's reconstruction performance without providing explicitly controllable security.
Furthermore, the design of the encryption key (e.g., key size, key distribution) remains unclear and lacks interpretability. 
Moreover, this method requires a secure channel for key distribution, which introduces practical challenges.
In summary, encryption-based methods impose a high communication load for delivering the secret keys, which significantly increases the cost compared to regular communication due to the security requirements.
Moreover, most of the aforementioned secure SemCom approaches are based on analog systems, 
while real-world communication systems are predominantly digital, presenting further challenges for practical deployment.


Building on our previous work \cite{chen2024nearly} and inspired by traditional coding schemes and encoded jammers (EJ) \cite{xu2024coding}, we propose a coding-enhanced jamming approach to achieve secure digital SemCom over wiretap channels, providing explicitly controllable system security. 
Specifically, we first encode the data in a private knowledge base using an NN-based encoder to generate a private codebook, where each codeword $\textbf{Y}_2(I)$ is associated with an index $I$. 
Second, we use another NN-based encoder to encode the transmitted data as $\textbf{Y}_1$. 
After that, we combine $\textbf{Y}_1$ with a randomly selected $\textbf{Y}_2(I)$ from the private codebook by superposition coding, where $\textbf{Y}_2$ plays the role of a jamming signal, and it is superposed as the inner code. 
The power allocation between the sequences is dynamically adjusted based on the symbol error probabilities (SEPs) of both the legitimate user and the eavesdropper.
Finally, the transmitter sends the superposed code over the regular channel, while sharing the index $I$ 
of the selected codeword $\textbf{Y}_2(I)$ 
with the legitimate receiver via a secure channel. 
This allows the legitimate receiver to decode the transmitted data using successive decoding.

The key contribution of our work is to take the idea from the previous paragraph into a practical secure SemCom system that leverages private knowledge to enhance security.
Specifically, our technical contributions are threefold.
First, we establish the general framework design of the proposed SemCom system. This framework includes multiple modules, such as the NN-based joint coding and modulation module, successive decoding module, and encoded jamming module.
%
Second, we propose a novel loss function for semantic coding, incorporating the normalized Hilbert-Schmidt independence criterion (nHSIC) as the regularization term. 
This term reduces the dependency between the jamming signal and the semantic information, preventing information leakage when the private knowledge and transmitted data are correlated.
Third, we develop a security analysis framework to guide the power allocation between $\textbf{Y}_1$ and $\textbf{Y}_2$. 
By dynamically adjusting the power allocation based on the derived SEPs, this framework explicitly controls system security, ensuring adaptable performance under varying channel conditions and security requirements.
%
Overall, compared to existing secure SemCom methods that rely on channel advantage or incorporate security solely as a training penalty term, the proposed method enables explicit and tunable control over system security without requiring any channel SNR gap. In addition, by eliminating the need for secret key transmission, it reduces communication overhead and improves practical deployability.
%
Numerical results demonstrate that even when the legitimate user and the eavesdropper experience identical channel conditions, the proposed method achieves security comparable to state-of-the-art approaches while improving the reconstruction performance of the legitimate user by over 1 dB under different channel SNRs and compression ratios.

\section{Related Work}


\subsection{Jammer-aided Secure Communications}


Jamming has been recognized as an effective method for enhancing the security of communication systems over wiretap channels \cite{huo2017jamming} and has gained widespread attention and application \cite{sun2022ris}. 
One of the most common jamming schemes involves transmitting Gaussian noise (GN), but this scheme simultaneously interferes with both the legitimate user and the eavesdropper. 
Consequently, specialized signal processing techniques are required to cancel the interference for the legitimate user.
To address this issue, Hu \textit{et al.} \cite{hu2017cooperative} proposed a cooperative jamming scheme to enhance the security of wireless networks.
In this scheme, the source node (Alice) and the cooperative jammer (Charlie) work together, with Alice transmitting a confidential message while Charlie generates Gaussian noise to help secure the transmission.
To prevent interference to the legitimate user (Bob), Charlie employs zero-forcing beamforming, ensuring that the noise only disrupts the eavesdroppers, while leaving Bob's communication unaffected.
%

A more advanced jamming scheme involves using an encoded jammer (EJ) to generate the jamming signal. 
Unlike the GN scheme, the EJ scheme selects appropriate codewords from a pre-designed codebook for transmission. Under certain conditions, this scheme allows the legitimate user to cancel the interference, while the eavesdropper is unable to do so, thereby ensuring the security of the source information.
The EJ scheme has been shown to achieve a higher secrecy rate compared to the GN scheme \cite{he2014providing}.
Xu \textit{et al.} \cite{xu2024coding} proposed a cooperative jamming scheme in which the jammer can dynamically switch between GN and EJ modes, significantly improving the secrecy performance of wireless communication systems. 
In the EJ mode, the jammer selects and transmits optimized codewords from the codebook based on the channel conditions. 
This ensures that the legitimate user can decode the jamming signal, cancel the interference, and recover the source information. 
The authors also proposed a low-complexity solution using the simultaneous diagonalization (SD) technique, which optimizes power allocation and precoder design to efficiently maximize the secrecy rate while reducing computational complexity.

\subsection{Secure Semantic Communications}

Regarding secure SemCom systems, Lin \textit{et al.} \cite{lin2023blockchain} proposed a blockchain-aided SemCom framework for AI-generated content in the metaverse. 
This scheme leverages blockchain and zero-knowledge proofs to ensure the security of semantic data against malicious tampering, while guaranteeing data integrity in decentralized systems.
Chen \textit{et al.} \cite{chen2023model} introduced the model inversion eavesdropping attack (MIEA) to highlight the privacy leakage risks in SemCom systems.
In this attack, adversaries attempt to reconstruct the source information by eavesdropping on transmitted signals and performing model inversion. The study addresses both white-box and black-box attack scenarios and presents a defense mechanism based on random permutation and substitution to mitigate MIEA attacks.
%
Li \textit{et al.} \cite{li2024secure} proposed a deep neural network-driven secure SemCom system (DeepSSC) based on physical layer security.
The training of DeepSSC is conducted in two phases to balance security and reliability. 
The first phase focuses on ensuring the reliability of SemCom, while the second phase aims to reduce the leakage of semantic information to eavesdroppers. 
Mu \textit{et al.} \cite{mu2024semantic} developed a physical layer security transmission framework based on SemCom, where semantic flows are used as artificial noise (AN) to interfere with malicious nodes and ensure information security.
Through optimization of power allocation and interference cancellation sequence, they maximized the secrecy rate over fading wiretap channels.
Qin \textit{et al.} \cite{qin2023securing} proposed a physical layer semantic encryption scheme to encrypt semantic data by exploring the randomness of bilingual evaluation understudy (BLEU) scores in machine translation. They also introduced a novel subcarrier-level semantic obfuscation mechanism to further strengthen communication security.


\section{Problem Setup and System Design}

\begin{figure}[t]
\begin{center}
\centerline{\includegraphics[width=1\linewidth]{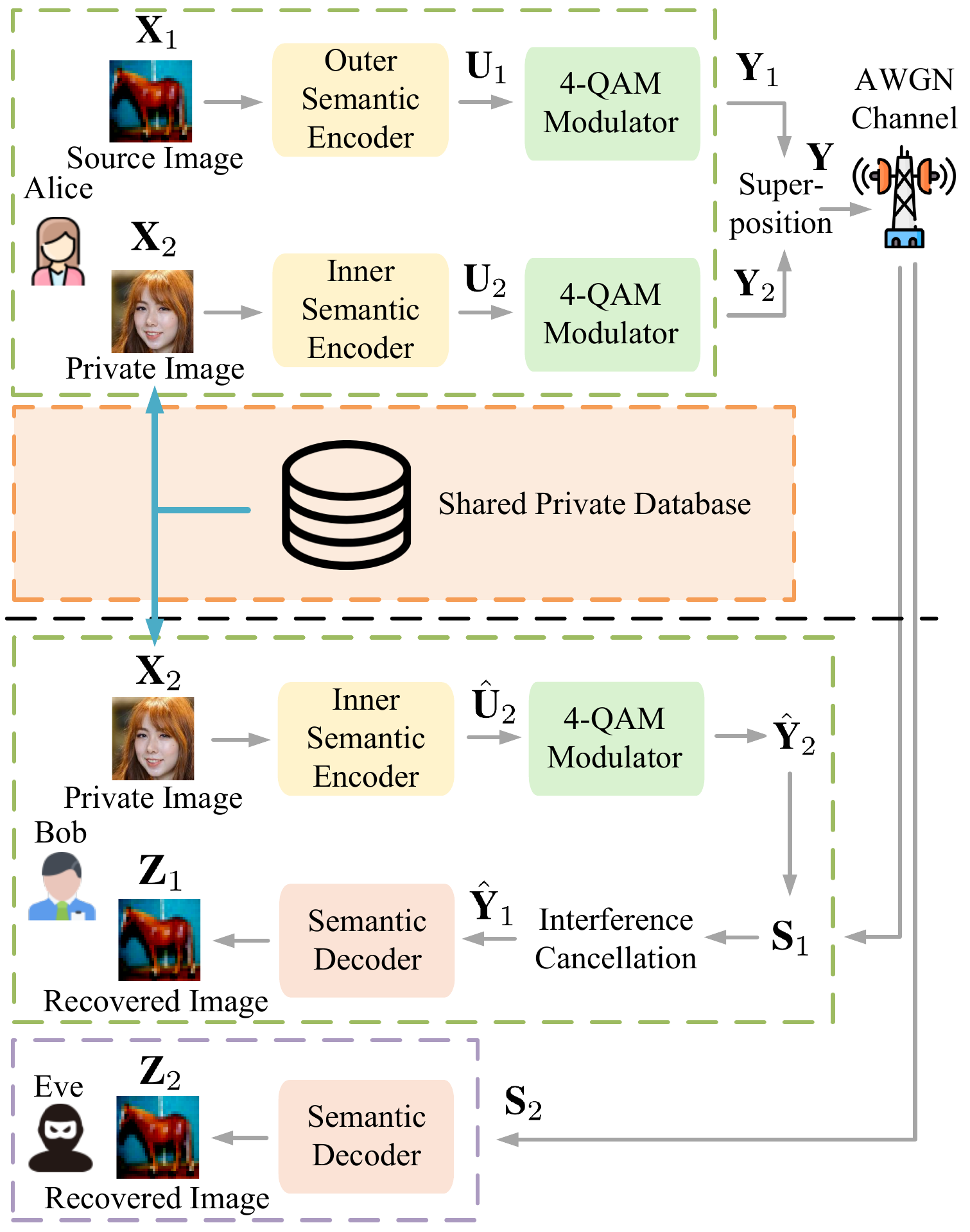}}
\caption{The proposed secure digital SemCom system with coding-enhanced jamming.}
\label{fig.1}
\end{center}
\vskip -0.3in
\end{figure}

\subsection{Problem Setup}



As shown in Fig.~\ref{fig.1}, we consider a secure digital SemCom system with an encoded jammer, designed for wireless image transmission over an additive white Gaussian noise (AWGN) wiretap channel. 
The transmitter, referred to as Alice, aims to transmit a source image to a legitimate receiver, Bob, as accurately as possible through the AWGN channel. Meanwhile, an eavesdropper, Eve, attempts to intercept the transmitted symbols via her own AWGN channel and recover the source image. 
%
%
Additionally, Alice and Bob share a private database of images. Before transmission, they exchange the index of an image from this private database, which serves as the private image used to generate the jamming signal and is unknown to Eve.
%
In this work, we assume that the index of the private image has been securely exchanged between Alice and Bob before transmission.
In practice, the index can be shared using a Diffie-Hellman key exchange protocol combined with authentication. Since the private database is fixed or infrequently updated, any updates can be transmitted using authenticated encryption.

We denote the source image Alice wants to transmit as $\textbf{X}_1$ and the private image known only to Alice and Bob as $\textbf{X}_2$. 
Alice first extracts the semantic information $\textbf{U}_1$ from the source image $\textbf{X}_1$ using an outer semantic encoder, denoted by
\begin{equation}
\textbf{U}_{1} = f_{\mathrm{ose}}\left( {\textbf{X}_1;\bm{\theta}^{\mathrm{ose}}} \right),
\end{equation}
where $f_{\mathrm{ose}}$ represents the outer semantic encoder.

The semantic information $\textbf{U}_{1}$ is then fed into an NN-based outer 4-QAM modulator to generate the outer constellation sequence $\textbf{Y}_{1}$, denoted by 
\begin{equation}
\textbf{Y}_{1} = f_{\mathrm{omod}}\left( {\textbf{U}_{1};\bm{\theta}^{\mathrm{omod}}} \right),
\end{equation}
where $f_{\mathrm{omod}}$ represents the outer 4-QAM modulator.

Alice also uses an inner semantic encoder to extract the semantic information $\textbf{U}_2$ from the private image $\textbf{X}_2$, denoted by
\begin{equation}
\textbf{U}_{2} = f_{\mathrm{ise1}}\left( {\textbf{X}_2;\bm{\theta}^{\mathrm{ise1}}} \right),
\end{equation}
where $f_{\mathrm{ise1}}$ represents the inner semantic encoder of Alice.

The semantic information $\textbf{U}_2$ is then fed into the inner 4-QAM modulator of Alice to generate the inner constellation sequence, i.e., the jamming signal $\textbf{Y}_{2}$, denoted by
\begin{equation}
\textbf{Y}_{2} = f_{\mathrm{imod1}}\left( {\textbf{U}_{2};\bm{\theta}^{\mathrm{imod1}}} \right),
\end{equation}
where $f_{\mathrm{imod1}}$ represents the inner 4-QAM modulator of Alice.

The lengths of $\textbf{Y}_{1}$ and $\textbf{Y}_{2}$ are equal, and both are normalized to satisfy the average power constraint $P$.
%
%
After obtaining $\textbf{Y}_{1}$ and $\textbf{Y}_{2}$, Alice scales and combines these two 4-QAM constellation sequences to form a 16-QAM constellation sequence $\textbf{Y}$. 
Alice then transmits $\textbf{Y}$ over the wiretap channel with AWGN. 
During the superposition operation, the power allocation between $\textbf{Y}_{1}$ and $\textbf{Y}_{2}$ is controlled by a power allocation coefficient (PAC) $a \in (0,0.5)$. 
Different PAC values correspond to varying levels of system security, which will be discussed later.
Our superposition operation can be mathematically expressed as
\begin{equation}
\textbf{Y} = \sqrt{a} \cdot \textbf{Y}_1 + \sqrt{1 - a} \cdot \textbf{Y}_2.
\end{equation}

$\textbf{Y}$ is subsequently transmitted to Bob over the AWGN channel.
Bob receives the noisy constellation sequence $\textbf{S}_{1}$, denoted by
\begin{equation}
  \textbf{S}_{1} = \textbf{Y} + \textbf{n}_1,   
\end{equation}
where $\textbf{n}_1 \sim \mathcal{CN} (0, {\sigma_{1}^{2}} )$.
Meanwhile, Eve eavesdrops on the transmitted constellation sequence through her own AWGN channel and obtains another noisy constellation sequence $\textbf{S}_{2}$, denoted by
\begin{equation}
  \textbf{S}_{2} = \textbf{Y} + \textbf{n}_2,   
\end{equation}
where $\textbf{n}_2 \sim \mathcal{CN} (0, {\sigma_{2}^{2}} )$.
Generally, we have $\sigma_1 \leq \sigma_2$.

The channel SNR between Alice and Bob/Eve is expressed as
\begin{equation}
    \mathrm{SNR}_{\rm leg}/\mathrm{SNR}_{\rm eve} = 10 \log_{10}\left(\frac{P}{\sigma^{2}_{1/2}}\right) \, (\mathrm{dB}).
\end{equation}

At the receiver, both Bob and Eve share the goal of reconstructing the source image $\textbf{X}_1$ as accurately as possible. 
For Bob, who has already exchanged the private image $\textbf{X}_2$ with Alice and knows the PAC value used during the superposition process,
he can attempt to cancel the interference present in $\textbf{S}_1$. 
Specifically, Bob first uses his own inner semantic encoder to extract semantic information $\hat{\textbf{U}}_2$ from the private image, denoted by
\begin{equation}
\hat{\textbf{U}}_{2} = f_{\mathrm{ise2}}\left( {\textbf{X}_2;\bm{\theta}^{\mathrm{ise2}}} \right),
\end{equation}
where $f_{\mathrm{ise2}}$ represents the inner semantic encoder of Bob.




Subsequently, $\hat{\textbf{U}}_{2}$ is fed into the inner 4-QAM modulator of Bob to generate an estimate of the inner constellation sequence, $\hat{\textbf{Y}}_{2}$, denoted by
\begin{equation}
\hat{\textbf{Y}}_{2} = f_{\mathrm{imod2}}\left( {\hat{\textbf{U}}_{2};\bm{\theta}^{\mathrm{imod2}}} \right),
\end{equation}
where $f_{\mathrm{imod2}}$ represents the inner 4-QAM modulator of Bob.
%
To ensure effective interference cancellation, $\hat{\textbf{Y}}_{2}$ should be as consistent as possible with $\textbf{Y}_2$.

After obtaining $\hat{\textbf{Y}}_{2}$, 
Bob performs interference cancellation based on $a$, $\textbf{S}_1$, and $\hat{\textbf{Y}}_{2}$, 
i.e., removing the interference of $\textbf{Y}_2$ from $\textbf{S}_1$ and recovering $\textbf{Y}_1$ as accurately as possible.
%
Specifically, our proposed interference cancellation process can be mathematically expressed as
\begin{equation}
\hat{\textbf{Y}}_1 = \frac{\textbf{S}_1 - \sqrt{1 - a} \cdot \hat{\textbf{Y}}_{2}}{\sqrt{a}},
\end{equation}
where $\hat{\textbf{Y}}_1$ is the estimate of the outer constellation sequence.


Bob then feeds $\hat{\textbf{Y}}_1$ into his semantic decoder to obtain the recovered image $\textbf{Z}_1$, denoted by
\begin{equation}
\textbf{Z}_1 = f_{\mathrm{sd1}}\left( {\hat{\textbf{Y}}_1;\bm{\theta}^{\mathrm{sd1}}} \right),
\end{equation}
where $\textbf{Z}_1$ represents Bob's recovered image, and $f_{\mathrm{sd1}}$ represents the semantic decoder of Bob.


For Eve, since she is unaware that the received constellation sequence is a superposition code and does not have access to the private image $\textbf{X}_2$, 
she directly inputs $\textbf{S}_2$ into her semantic decoder to obtain the recovered image $\textbf{Z}_2$, denoted by
\begin{equation}
\textbf{Z}_2 = f_{\mathrm{sd2}}\left( {\textbf{S}_2;\bm{\theta}^{\mathrm{sd2}}} \right),
\end{equation}
where $\textbf{Z}_2$ represents Eve's recovered image, and $f_{\mathrm{sd2}}$ represents the semantic decoder of Eve.


%
The goal of our proposed system is to maximize the image reconstruction performance of the legitimate user while ensuring that the image reconstruction performance of the eavesdropper does not exceed a predefined constraint.

We evaluate the image reconstruction performance of the users using the peak signal-to-noise ratio (PSNR) metric, which we refer to as PSNR performance. 
%
For instance, the PSNR performance of the legitimate user is calculated using the following equation:
\begin{equation}
\text{PSNR}(\textbf{X}_1,\textbf{Z}_1) = 10\log_{10}\left( \frac{{\text{MAX}}^{2}}{\text{MSE}(\textbf{X}_1,\textbf{Z}_1)} \right) (\mathrm{dB}),
\end{equation}
where $\text{MAX}$ denotes the highest pixel value in the source image, which is 255 for a 24-bit RGB image.
The $\text{MSE}$ represents the mean squared error.

\subsection{Proposed Secure SemCom System}


%
%
Our proposed system consists of four key components, including the outer semantic encoder, inner semantic encoder, NN-based 4-QAM modulator, and semantic decoder.  
Building on our previous work \cite{bo2024deep,chen2024nearly}, we design the network architectures for the semantic encoder, 4-QAM modulator, and semantic decoder.
Additionally, the generation of the inner constellation sequence and the design of the superposition constellation map are crucial elements of the system design.
%
The inner constellation sequence is obtained by encoding and modulating the private image. 
Since this constellation sequence acts as a jamming signal, it must be distinct from the outer constellation sequence.
This process involves two critical aspects, selecting the private image and ensuring that the inner constellation sequence remains different from the outer one.
We generate a superposition constellation sequence by combining the outer and inner constellation sequences to reduce the information leakage to the eavesdropper. 
This process directly corresponds to system security, which will be further discussed in Section IV.
The following sections provide a detailed explanation of each component.

\subsubsection{Outer and Inner Semantic Encoders}


\begin{figure}[t]
\begin{center}
\centerline{\includegraphics[width=1\linewidth]{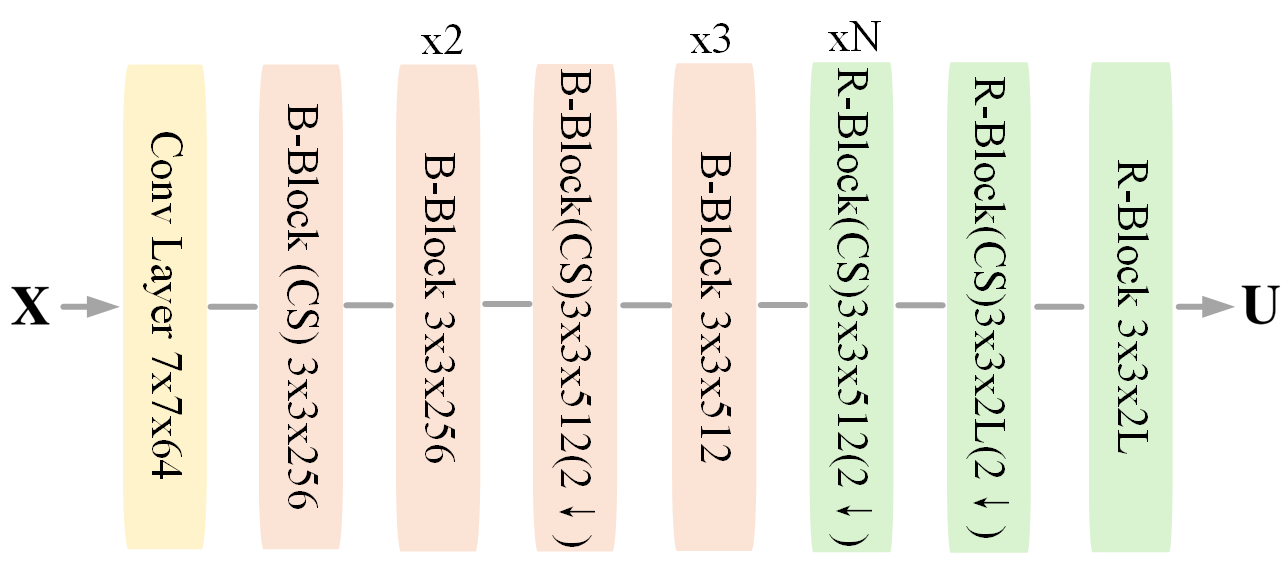}}
\caption{The network architecture of the outer and inner semantic encoders.}
\label{se}
\end{center}
\vskip -0.3in
\end{figure}

The network architectures of the outer and inner semantic encoders are nearly identical, with the primary difference being the number of layers in the neural network.
As the size of the input image increases, more layers are required in the semantic encoder.
%
The semantic encoder extracts the semantic information $\textbf{U}$ from the input image $\textbf{X} \in \mathbb{R}^{H \times W \times C}$.
%
The network architecture of the outer and inner semantic encoders is shown in Fig.~\ref{se}. 
%
The semantic encoder consists of a single convolutional layer, seven bottleneck blocks (B-Blocks), and $N+2$ resnet blocks (R-Blocks), where $ S_{\rm k} \times S_{\rm k} \times C_{\rm o}$ below each layer or bottleneck/resnet block represents its configuration.
$C_{\rm o}$ represents the number of output channels of the convolutional layer or the bottleneck/resnet block. 
For the convolutional layer, $S_{\rm k}$ denotes its kernel size.
For a bottleneck/resnet block, $S_{\rm k}$ indicates the maximum kernel size of the convolutional layers along its main path.
The $ 2 \downarrow $ symbol indicates down-sampling with a stride of 2, and ``Conv Shortcut (CS)'' indicates that a convolutional layer is present in the shortcut connection of the block. 
%
The depth of the semantic encoder is controlled by the variable $N$.
The output of the semantic encoder is the semantic information $\textbf{U} \in \mathbb{R}^{\frac{H}{2^{2+N}} \times \frac{W}{2^{2+N}} \times 2L}$,
where $2+N$ represents the number of down-sampling operations performed by the semantic encoder, 
and $L$ controls the length of $\textbf{U}$.



\subsubsection{4-QAM Modulator}

%
Since our proposed system is a digital SemCom system, it is necessary to convert the analog semantic information symbols into discrete constellation sequences. 
Directly quantizing analog symbols into digital signals involves discrete sampling, which is non-differentiable in deep learning. 
This leads to the vanishing gradient problem, making it difficult to train the system.
To address this issue, we use an NN-based 4-QAM modulator, as proposed in \cite{bo2022learning}, 
to learn how to map the analog semantic information symbols $\textbf{U}$ to the discrete 4-QAM constellation sequence $\textbf{Y}$.
By applying the Gumbel-Softmax sampling method \cite{jang2016categorical} to sample the analog symbols, this approach successfully avoids the non-differentiability issue caused by discrete sampling. 
Next, we provide a detailed description of the 4-QAM modulator design.

\begin{figure}[t]
\begin{center}
\centerline{\includegraphics[width=0.67\linewidth]{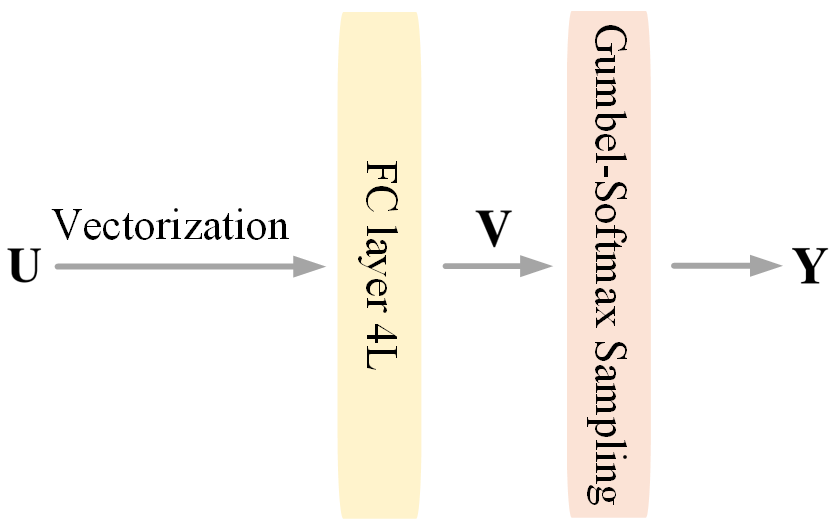}}
\caption{The network architecture of the NN-based 4-QAM modulator.}
\label{fig.4qam}
\end{center}
\vskip -0.3in
\end{figure}


The network architecture of the outer and inner 4-QAM modulators is identical, as shown in Fig.~\ref{fig.4qam}. 
The analog semantic information $\textbf{U}$ is input into the NN-based 4-QAM modulator, first being vectorized into $\bar{\textbf{U}}$, and then passed through a fully connected (FC) layer to produce the output vector $\textbf{V}$. 
The number $4L$ under the FC layer represents the number of output neurons, meaning the length of $\textbf{V}$ is $4L$. 
$\textbf{V}$ is a probability distribution sequence consisting of 
$L$ vectors 
$\textbf{V}_{i(i=1,2, ..., L)}=\left( P_{i}^{1}, P_{i}^{2}, P_{i}^{3}, P_{i}^{4} \right)$, each of length 4. 
The four elements in each vector $\textbf{V}_i$ represent the probabilities of selecting the $i$-th 4-QAM constellation symbol as $\left(1+1j, 1-1j, -1+1j, -1-1j\right)$, respectively. 
Therefore, this FC layer learns the probability distribution of the 4-QAM constellation sequence $\textbf{Y}$.
Subsequently, we use the Gumbel-Softmax sampling method \cite{jang2016categorical} to sample from $\textbf{V}$, 
generating the discrete 4-QAM constellation sequence $\textbf{Y} \in \mathbb{C}^{L \times 1}$. 
The length of $\textbf{Y}$ is $L$, with each element being a complex symbol. 
The Gumbel-Softmax sampling method transforms this discrete operation into a differentiable process, thus avoiding the non-differentiability issue in discrete sampling. 
This allows our proposed system to perform successful backpropagation during training.

\subsubsection{Generation of the Inner Constellation Sequence}


The inner constellation sequence $\textbf{Y}_2 \in \mathbb{C}^{L \times 1}$ is generated from the private image $\textbf{X}_2$ through the inner semantic encoder and 4-QAM modulator. 
While $\textbf{Y}_2$ carries no valid information, it serves as a jamming signal designed to enhance system security. 
To ensure the effectiveness of this jamming signal against eavesdropping, the difference between $\textbf{Y}_2$ and the outer constellation sequence $\textbf{Y}_1 \in \mathbb{C}^{L \times 1}$ must be sufficiently large. 
Thus, when selecting the private image $\textbf{X}_2$, we ensure that it significantly differs from the source image $\textbf{X}_1$ in terms of its category.
For example, if the source image is of an animal, we may choose a facial image as the private image.
Additionally, the size of the private image need not match that of the source image, as the depth of the inner semantic encoder is adjustable.

In addition to ensuring the effectiveness of the jamming signal through careful selection of the private image, we also focus on distinguishing $\textbf{Y}_1$ and $\textbf{Y}_2$ from a signal encoding and modulation perspective.
Specifically, our objective is to minimize the MI between $\textbf{Y}_1$ and $\textbf{Y}_2$ during system training. 
In other words, we aim to make $\textbf{Y}_1$ and $\textbf{Y}_2$ as independent as possible, thereby enhancing the interference effect of $\textbf{Y}_2$. 
However, due to the complexity of approximating the distribution of hidden representations in convolutional neural networks, directly computing the MI between $\textbf{Y}_1$ and $\textbf{Y}_2$ is challenging. 
To address this issue, we adopt the method proposed in \cite{zheng2021information}, which uses the normalized Hilbert-Schmidt independence criterion (nHSIC) as a replacement for MI.
The nHSIC between two random variables $\textbf{X}$ and $\textbf{Y}$ is defined as:
\begin{equation}
    \text{nHSIC}(\textbf{X}, \textbf{Y}) = \frac{\text{tr}(\textbf{K}_\textbf{X} \textbf{K}_\textbf{Y})}{\sqrt{\text{tr}(\textbf{K}_\textbf{X} \textbf{K}_\textbf{X})}\sqrt{ \text{tr}(\textbf{K}_\textbf{Y} \textbf{K}_\textbf{Y})}},
\end{equation}
where $\textbf{K}_\textbf{X}$ is the kernel matrix for $\textbf{X}$, and each element of the matrix $\textbf{K}_{\textbf{X}(i,j)} $ is $k(x_i, x_j)$, computed using the kernel function $k$. 
Here, $x_i$ and $x_j$  represent two samples from $\textbf{X}$. It has been shown in \cite{zheng2021information} that when using the linear kernel, minimizing nHSIC is equivalent to minimizing the MI under the Gaussian assumption. Therefore, we can minimize $\text{nHSIC}(\textbf{Y}_1, \textbf{Y}_2)$ during system training to reduce the similarity between $\textbf{Y}_1$ and $\textbf{Y}_2$.

\subsubsection{Design of the Superposition Constellation Map}

%
The design of the superposition constellation map is directly linked to system security. 
%
Since the jamming signal contains no valid information, both the legitimate user and the eavesdropper aim to decode the outer 4-QAM constellation symbols from the received semantic information.
At the receiver, the legitimate user can remove the interference from the received semantic information using the private image. 
As a result, the legitimate user essentially receives a 4-QAM constellation sequence and decodes the outer 4-QAM constellation symbols based on this signal.
On the other hand, the eavesdropper cannot cancel the interference.
Therefore, the eavesdropper attempts to decode the outer 4-QAM constellation symbols from the received 16-QAM superposition constellation sequence.
This results in a scenario where, although both users may have the same channel SNR, they experience different SEPs when decoding the outer 4-QAM constellation symbols. 
The SEP is controlled by the PAC $a$, and varying PAC values lead to different SEPs for both the legitimate user and the eavesdropper.
%
Therefore, 
we can control the system's security and performance by adjusting the PAC. 
The detailed design of the superposition constellation map will be described in Section IV.

\subsubsection{Semantic Decoder}

\begin{figure}[t]
\begin{center}
\centerline{\includegraphics[width=0.91\linewidth]{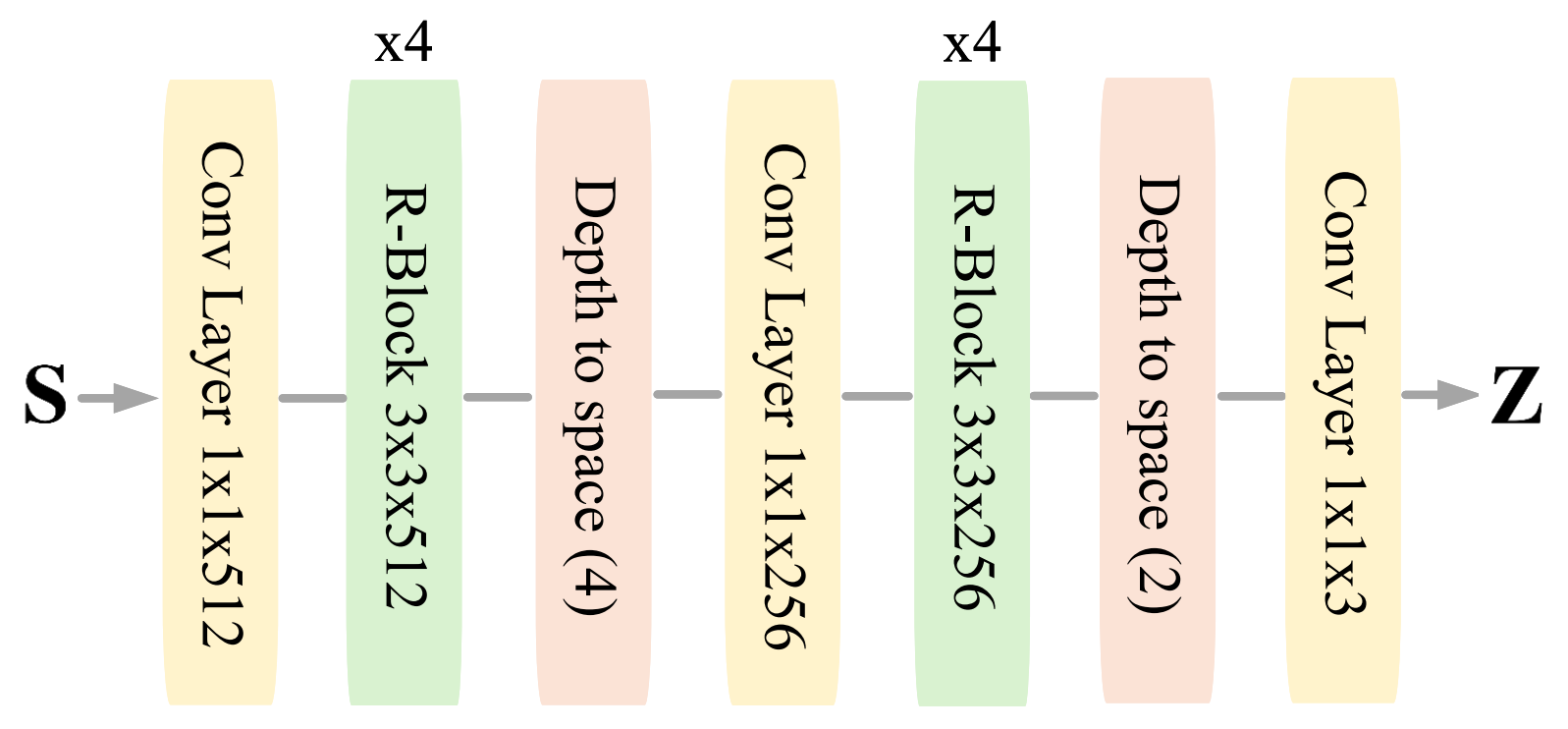}}
\caption{The network architecture of the semantic decoder.}
\label{fig.sd}
\end{center}
\vskip -0.3in
\end{figure}


We assume that the eavesdropper has stolen the network architecture and parameters of the legitimate user's semantic decoder, which is a challenging assumption in secure SemCom research. 
As a result, the network architecture of the eavesdropper's semantic decoder is identical to that of the legitimate user's, as shown in Fig.~\ref{fig.sd}. 
The semantic decoder attempts to reconstruct the source image based on the received semantic information $\textbf{S}$, in order to obtain the recovered image $\textbf{Z}$. 
The semantic decoder consists of three convolutional layers, eight resnet blocks, and two \textit{depth to space} blocks.
The convolutional layers and resnet blocks are primarily responsible for processing feature maps and do not perform up-sampling.
%
The \textit{depth to space} blocks, on the other hand, are responsible for up-sampling the feature maps.
The number $s$ following the \textit{depth to space} block indicates its configuration. 
Specifically, after passing through the \textit{depth to space} block, the number of channels in the feature map is reduced by a factor of  $2^{s}$, 
while its height and width are increased by a factor of $s$. 
Finally, the output of the semantic decoder is the recovered image $\textbf{Z} \in \mathbb{R}^{H \times W \times C}$.

\subsection{Training Strategy}


In this paper, we propose a three-stage training strategy for the secure SemCom system introduced above. 
By leveraging this multi-stage training strategy, we can simulate a strong eavesdropper and effectively optimize system security accordingly.
We assume that during the actual transmission of the source image (after training), the eavesdropper steals the network architecture and parameters of the legitimate user's semantic decoder to reconstruct the source image.
However, the eavesdropper cannot perform interference cancellation. 
Additionally, we assume that the eavesdropper and the legitimate user have identical channel SNRs, which represents the most favorable scenario for the eavesdropper and allows us to fully demonstrate the effectiveness of our proposed jamming scheme.
As a result, the performance gap between the eavesdropper and the legitimate user primarily depends on the effectiveness of the jamming signal.


In the first training stage, we train only Alice's outer and inner semantic encoders, outer and inner 4-QAM modulators, and Bob's semantic decoder. 
During this stage, we assume that $\hat{\textbf{Y}}_2$ and $\textbf{Y}_2$ are identical, and therefore, we do not train Bob's inner semantic encoder and 4-QAM modulator. 
Additionally, the eavesdropper is not required in this stage. 
To enhance the effectiveness of the jamming signal, we minimize the similarity between $\textbf{Y}_1$ and $\textbf{Y}_2$.
The loss function of the first training stage is represented as
\begin{equation}
\mathcal{L}_{1}= \text{MSE}(\textbf{X}_1,\textbf{Z}_{1})+\lambda_1 \cdot \text{nHSIC}(\textbf{Y}_1, \textbf{Y}_2),
\end{equation}
where $ \text{MSE}(\textbf{X}_1,\textbf{Z}_{1})$ measures the quality of Bob's recovered image, and $\text{nHSIC}(\textbf{Y}_1, \textbf{Y}_2)$ measures the similarity between $\textbf{Y}_1$ and $\textbf{Y}_2$.


In the second training stage, we train only Eve's semantic decoder. During this stage, we essentially train a simulated adversary Eve for Alice and Bob, so that in the third training stage, Alice and Bob can optimize their networks against this simulated adversary. 
This ensures the effectiveness of our jamming scheme and enhances system security. 
The loss function of the second training stage is represented as
\begin{equation}
\mathcal{L}_{2}= \text{MSE}(\textbf{X}_1,\textbf{Z}_{2}),
\end{equation}
where $ \text{MSE}(\textbf{X}_1,\textbf{Z}_{2})$ measures the quality of Eve's recovered image.


In the third training stage, we train all modules except Eve's semantic decoder. During this stage, Alice and Bob optimize their networks against Eve to improve Bob's reconstruction performance and reduce Eve's reconstruction performance. 
In addition to minimizing the similarity between $\textbf{Y}_1$ and $\textbf{Y}_2$, we also minimize the MSE between $\hat{\textbf{Y}}_2$ and $\textbf{Y}_2$ to ensure the effectiveness of interference cancellation. 
The loss function of the third training stage is represented as
\begin{equation}
\begin{split}
\mathcal{L}_{3}= \text{MSE}(\textbf{X}_1,\textbf{Z}_{1})+\lambda_1 \cdot \text{nHSIC}(\textbf{Y}_1, \textbf{Y}_2)+ \\
\lambda_2 \cdot\text{MSE}(\textbf{Y}_2,\hat{\textbf{Y}}_2)-\lambda_3 \cdot \text{MSE}(\textbf{X}_1,\textbf{Z}_{2}),
\end{split}
\end{equation}
where $\text{MSE}(\textbf{Y}_2,\hat{\textbf{Y}}_2)$ measures the accuracy of Bob's interference cancellation.
We determine $\lambda_1$, $\lambda_2$, and $\lambda_3$ through extensive trials to balance the trade-offs: higher $\lambda_1$ promotes jamming independence, higher $\lambda_2$ improves interference cancellation, and higher $\lambda_3$ more effectively suppresses Eve's reconstruction, all of which also impact the task performance of the legitimate user.


\section{Constellation Map Design}


The outer and inner 4-QAM constellation sequences 
$\textbf{Y}_1$  and $\textbf{Y}_2$  are combined into a 16-QAM constellation sequence $\textbf{Y} = \sqrt{a} \cdot \textbf{Y}_1 + \sqrt{1 - a} \cdot \textbf{Y}_2$ based on the PAC $a$, 
where $a \in (0,0.5)$. 
The values $a$ and $1 - a$ represent the power allocated to $\textbf{Y}_1$  and $\textbf{Y}_2$, respectively, with a total power of 1. 
In this section, we will discuss the SEP of both the legitimate user and the eavesdropper when decoding the outer constellation symbols.
These two SEPs are directly linked to the task performance and security of our proposed system.
Therefore, we will also analyze the setting of the PAC based on these SEPs.


\subsection{The SEP of the Legitimate User}


\begin{figure}[t]
\begin{center}
\centerline{\includegraphics[width=0.45\linewidth]{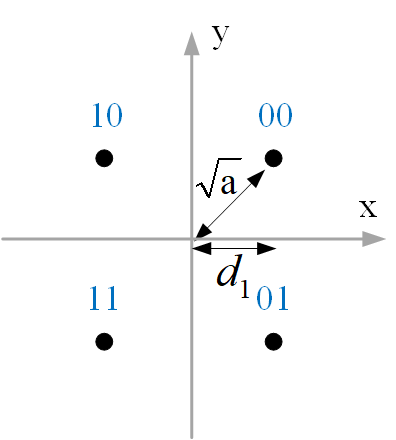}}
\caption{The 4-QAM outer constellation map.}
\label{fig.outer}
\end{center}
\vskip -0.3in
\end{figure}

Assuming that the interference cancellation at the receiver for the legitimate user is sufficiently effective, the transmitted signal can be considered as the outer constellation sequence $\textbf{Y}_1$, 
while the received signal can be considered as $\textbf{Y}_1$ corrupted by AWGN noise, where the standard deviation of the AWGN noise is $\sigma$. 
The noise follows a Gaussian distribution with the probability density function $\mathcal{N}(x;0,\sigma^2)$. 
The noise components are independent along the x-axis and y-axis.
The SEP of the legitimate user when decoding the outer constellation symbols depends on $\sigma$ and the PAC $a$. 
The outer constellation map $\textbf{Y}_1$ is shown in Fig.~\ref{fig.outer}, where Alice transmits each symbol with approximately equal probability.
The distance of the outer constellation points from the origin is $\sqrt{a}$, and we have $d_{1}= \sqrt{a/2}$. 
We first calculate the symbol correctness probability (SCP) of the legitimate user when decoding the outer constellation symbols using  the maximum likelihood (ML) detector, and then derive the SEP.


Let $SCP_{s_2}^{s_1}$ represent the probability that the received constellation point falls in region $s_2$ when the transmitted symbol is $s_1$. 
Suppose the transmitted symbol is ``00'', and we define its location as the origin for the maximum likelihood decision rule. 
If the received constellation point falls within the region ``00'', the decoding is considered correct.
Based on this, we calculate $SCP_{00}^{00}$ as follows:
\begin{equation}
\begin{split}
 & SCP_{00}^{00} = \int_{-d_1}^{\infty} \int_{-d_1}^{\infty} f_{X,Y}(x, y) \, dx \, dy \\
& =\int_{-d_1}^{\infty}\frac{1}{\sqrt{2\pi}}exp\left(-\frac{\left(\frac{x}{\sigma}\right)^2}{2}\right)d\left(\frac{x}{\sigma}\right) \cdot \\
& \int_{-d_1}^{\infty}\frac{1}{\sqrt{2\pi}}exp\left(-\frac{\left(\frac{y}{\sigma}\right)^2}{2}\right)d\left(\frac{y}{\sigma}\right) \\ 
& =  {Q(\frac{{-d}_1}{\sigma})}^{2},
 \label{eq21}
 \end{split}
\end{equation}
where $f_{X,Y}(x, y)$ is the joint probability density function of the Gaussian noise. 
This is the product of the two marginal probability density functions, i.e., $f_{X,Y}(x,y) = f_X(x) \cdot f_Y(y) = \mathcal{N}(x;0,\sigma^2) \cdot \mathcal{N}(y;0,\sigma^2)$. 
Additionally, $Q(x)=\frac{1}{\sqrt{2\pi}}\int_{x}^{\infty}{exp(-\frac{u^2}{2})du}$, and $\sigma$ represents the standard deviation of the Gaussian noise.


Following the same procedure, we calculate $SCP_{01}^{01}$, $SCP_{10}^{10}$, and $SCP_{11}^{11}$.
Consequently, we find that
\begin{equation}
SCP_{00}^{00}=SCP_{01}^{01}=SCP_{10}^{10}=SCP_{11}^{11}.
\end{equation}
When the transmitted symbol is ``00'', 
the SEP of the legitimate user when decoding the outer constellation symbol ``00'' is 
\begin{equation}
SEP^{00}=1-SCP_{00}^{00}.
\label{eq23}
\end{equation}
%
Therefore, the SEP of the legitimate user when decoding the outer constellation symbols is the same for all four transmitted symbols. This can be expressed as
\begin{equation}
SEP^{00}=SEP^{01}=SEP^{10}=SEP^{11}.
\end{equation}
Thus, the overall SEP of the legitimate user when decoding the outer constellation symbols is
\begin{equation}
SEP=\frac{(SEP^{00}+SEP^{01}+SEP^{10}+SEP^{11})}{4}=SEP^{00}, 
\end{equation}
\label{SEP-calculate1}
where $SEP^{00}$ is given by equation (\ref{eq21}) and (\ref{eq23}).

\subsection{The SEP of the Eavesdropper}

\begin{figure}[t]
\begin{center}
\centerline{\includegraphics[width=0.82\linewidth]{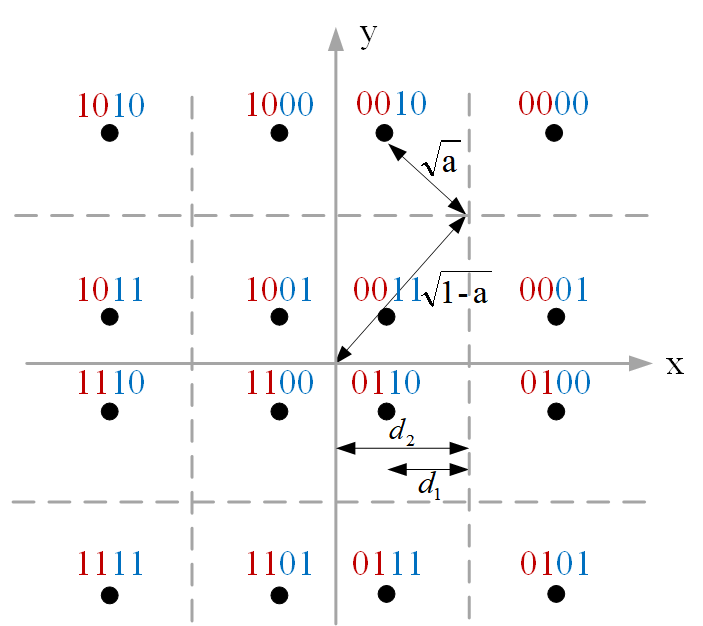}}
\caption{The 4-QAM+4-QAM superposition constellation map.}
\label{fig.superposition}
\end{center}
\vskip -0.3in
\end{figure}


For the eavesdropper, she cannot perform interference cancellation.
The transmitted signal is the superposition constellation sequence $\textbf{Y}$, while the received signal is $\textbf{Y}$ corrupted by AWGN noise. 
The relevant settings have been introduced in the previous subsection.
The superposition constellation map $\textbf{Y}$ is shown in Fig.~\ref{fig.superposition}, where the blue symbols represent the outer constellation and the red symbols represent the inner constellation.
Alice transmits each inner/outer constellation symbol with approximately equal probability.
The distance of the inner constellation points from the origin is $\sqrt{1-a}$, and the distance of the outer constellation points from the inner constellation points is $\sqrt{a}$, with $d_{1}= \sqrt{a/2}$, $d_{2}= \sqrt{(1-a)/2}$, and $d_{2}>d_{1}$. 


The eavesdropper aims to correctly decode the outer constellation symbols, as decoding the inner constellation symbols is irrelevant since they carry no valid information.
For example, when the transmitted inner constellation symbol is ``00'' and the transmitted outer constellation symbol is ``10'', 
the eavesdropper is considered to have correctly decoded the outer constellation symbol ``10'' if the received constellation point falls within any of the four regions: ``0010'', ``1010'', ``1110'', or ``0110''. 
Therefore, when the transmitted symbol is ``0010'', the SCP of the eavesdropper when decoding the outer constellation symbol ``10'' is given by
\begin{equation}
{SCP}_{10}^{0010}={SCP}_{0010}^{0010}+{SCP}_{1010}^{0010}+{SCP}_{1110}^{0010}+{SCP}_{0110}^{0010}.
\end{equation}

From Fig.~\ref{fig.superposition}, these SCPs can be derived as
\begin{equation}
\begin{split}
& SCP_{0010}^{0010} = \left[ Q(\frac{-(d_2-d_1)}{\sigma})-Q(\frac{d_1}{\sigma}) \right] \cdot Q(\frac{{-d}_1}{\sigma}). \\
& SCP_{1010}^{0010} = Q(\frac{2d_2-d_1}{\sigma}) \cdot Q(\frac{{-d}_1}{\sigma}). \\
& SCP_{1110}^{0010} = Q(\frac{2d_2-d_1}{\sigma}) \\
& \cdot \left[ Q(\frac{-(d_1+2d_2)}{\sigma}) - Q(\frac{-(d_1+d_2)}{\sigma})\right]. \\
& SCP_{0110}^{0010} = \left[Q(\frac{-(d_2-d_1)}{\sigma})-Q(\frac{d_1}{\sigma}) \right] \cdot \\ 
& \left[ Q(\frac{-(d_1+2d_2)}{\sigma}) - Q(\frac{-(d_1+d_2)}{\sigma})\right]. \\
\end{split}
\label{eq27}
\end{equation}

Following the same process, we can calculate the SCPs of the eavesdropper when decoding the outer constellation symbols ``00'', ``11'', and ``01'' for transmitted symbols ``0000'', ``0011'', and ``0001'', respectively. 
The calculations of ${SCP}_{00}^{0000}$, ${SCP}_{11}^{0011}$, and ${SCP}_{01}^{0001}$ follow the same steps as equation (\ref{eq27}) and are omitted for brevity.
Thus, when the transmitted inner constellation symbol is ``00'', 
the SEP of the eavesdropper when decoding the outer constellation symbols is 
\begin{equation}
SEP^{00}=1-\frac{({SCP}_{10}^{0010}+{SCP}_{00}^{0000}+{SCP}_{11}^{0011}+{SCP}_{01}^{0001})}{4}. 
\label{eq28}
\end{equation}

We then calculate the SEP of the eavesdropper when decoding the outer constellation symbols for the transmitted inner constellation symbols  
``01'', ``10'', and ``11''.
Note that the SCPs exhibit symmetry for different inner constellation symbols.
For instance, the SCP of the eavesdropper when decoding the outer constellation symbol ``00'' with the transmitted inner constellation symbol ``01'' is the same as the SCP of the eavesdropper when decoding the outer constellation symbol ``10'' with the transmitted inner constellation symbol ``00'', i.e., ${SCP}_{00}^{0100}={SCP}_{10}^{0010}$.
%
This symmetry continues for all four inner constellation symbols, meaning the SEP of the eavesdropper when decoding the outer constellation symbols is the same for all four inner constellation symbols.
Omitting the intermediate steps, we have
\begin{equation}
SEP^{00}=SEP^{01}=SEP^{10}=SEP^{11}.
\end{equation}

Finally, the overall SEP of the eavesdropper when decoding the outer constellation symbols is 
\begin{equation}
SEP=\frac{(SEP^{00}+SEP^{01}+SEP^{10}+SEP^{11})}{4}=SEP^{00}, 
\end{equation}
\label{SEP-calculate2}
where $SEP^{00}$ is given by equation (\ref{eq28}).

\begin{figure}[t]
\begin{center}
\centerline{\includegraphics[width=0.95\linewidth]{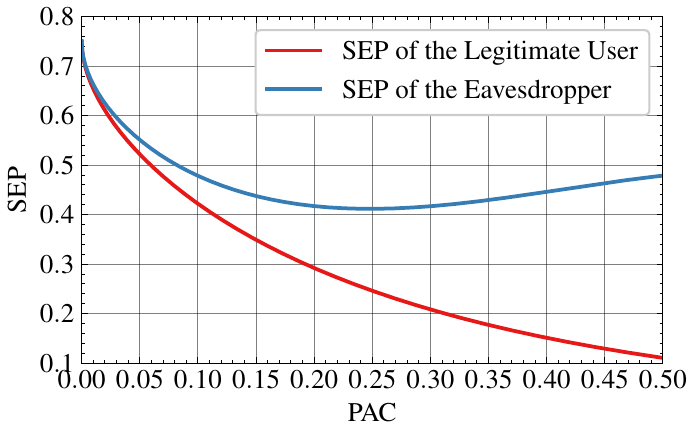}}
\caption{The SEP curves of both the legitimate user and the eavesdropper when decoding the outer constellation symbols. We have $\mathrm{SNR}_{\mathrm{leg}}=\mathrm{SNR}_{\mathrm{eve}}=10\mathrm{dB}$.}
\label{fig.seps}
\end{center}
\vskip -0.3in
\end{figure}

\subsection{The Setting of the Power Allocation Coefficient}


The channel SNRs of the legitimate user and the eavesdropper are denoted as $\mathrm{SNR}_{\rm leg}$ and $\mathrm{SNR}_{\rm eve}$, respectively. 
In this paper, we assume that the eavesdropper experiences the strongest channel conditions, identical to those of the legitimate user, to demonstrate the security of our proposed jamming scheme.
We plot the SEP curves of both the legitimate user and the eavesdropper as a function of the PAC $a$ for the scenario where $\mathrm{SNR}_{\rm leg}=\mathrm{SNR}_{\rm eve}=10\rm dB$, as shown in Fig.~\ref{fig.seps}. 
The red curve represents the SEP of the legitimate user, while the blue curve represents the SEP of the eavesdropper.
We observe that the SEP of the legitimate user decreases rapidly as $a$ increases, 
whereas the SEP of the eavesdropper initially decreases and then increases with $a$. 
Due to the jamming signal, the SEP of the eavesdropper is consistently higher than that of the legitimate user.
The SEP of the eavesdropper is directly related to system security, and based on the derived SEP curves, 
we can adjust the PAC value to explicitly control system security while ensuring high task performance.
For example, when $a=0.49$, 
the SEP of the legitimate user is low at 11.33\%, 
while the SEP of the eavesdropper is high at 47.66\%. 
This power allocation scheme achieves excellent task performance while maintaining high system security. 
On the other hand, when $a=0.40$, 
the SEP of the legitimate user is 15.14\%, and the SEP of the eavesdropper is 44.63\%, indicating a decrease in both security and task performance of the system. 
Thus, by adjusting the PAC values based on the derived SEP curves, we can flexibly achieve different levels of system security and task performance.

\begin{figure}[t]
\centering
\subfigure[$a=0.49$]{
\begin{minipage}[t]{1\linewidth}
\centering
\includegraphics[width=0.95\linewidth]{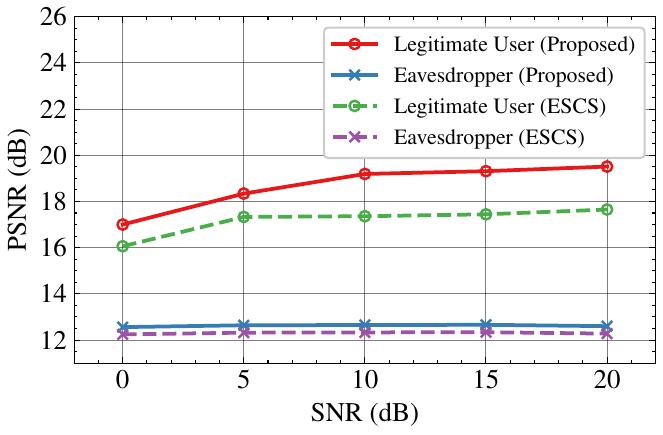}
\end{minipage}
}
\subfigure[$a=0.40$]{
\begin{minipage}[t]{1\linewidth}
\centering
\includegraphics[width=0.95\linewidth]{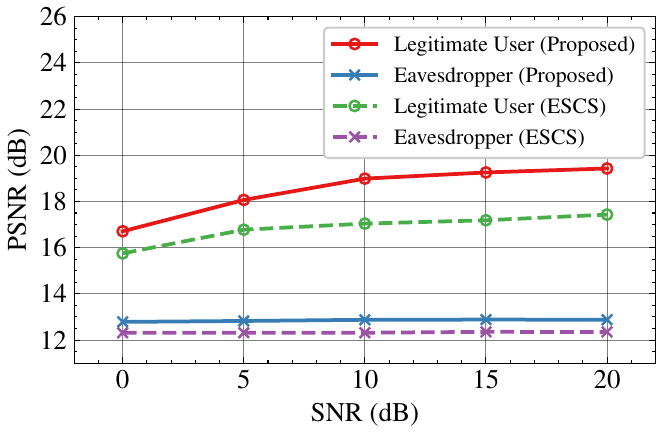}
\end{minipage}
}
\caption{The PSNR performance of our proposed method and ESCS for both users under different channel SNRs, with PAC values of 0.49 and 0.40. The private dataset is FFHQ-64, the compression ratio is 1/24, and $\rm SNR_{\rm leg}=\rm SNR_{\rm eve}$.
}
\label{fig.plot1}
\vskip -0.2in
\end{figure}

\begin{figure}[t]
\centering
\subfigure[$a=0.49$]{
\begin{minipage}[t]{1\linewidth}
\centering
\includegraphics[width=0.95\linewidth]{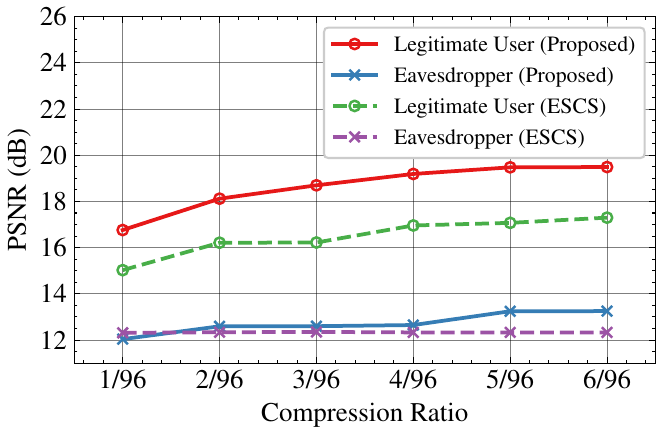}
\end{minipage}
}
\subfigure[$a=0.40$]{
\begin{minipage}[t]{1\linewidth}
\centering
\includegraphics[width=0.95\linewidth]{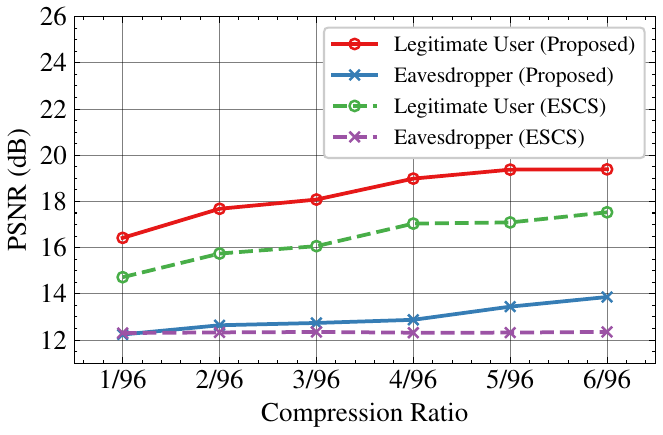}
\end{minipage}
}
\caption{The PSNR performance of our proposed method and ESCS for both users under different compression ratios, with PAC values of 0.49 and 0.40. The private dataset is FFHQ-64, the channel SNR is 10 dB, and $\rm SNR_{\rm leg}=\rm SNR_{\rm eve}$.
}
\label{fig.plot2}
\vskip -0.2in
\end{figure}

\section{Simulation Results}

\subsection{Experimental Settings}

\subsubsection{Datasets}



In our experiments, we use the CIFAR-10 and CIFAR-100 datasets for source image transmission, both consisting of $32 \times 32$ RGB images. 
The training sets contain 50,000 images, and the test sets contain 10,000 images.
We select two different private datasets to evaluate the impact of various types and complexities of private images on system security. 
The first private dataset is FFHQ-64 \cite{karras2019style}, a facial image dataset, where all images are $64 \times 64$ RGB images.
%
The training set consists of 60,000 images, from which we use 50,000 for training. The test set contains 10,000 images.
The second private dataset is MNIST \cite{lecun1998gradient}, consisting of $28 \times 28$ grayscale images.
It also contains 60,000 training images and 10,000 test images, from which we use 50,000 training samples. No data augmentation is applied to any of the datasets.

\subsubsection{Compression Ratios}


Assume that the number of symbols in the source image is $n$, and the number of real-valued symbols transmitted over the channel is $k$. 
The compression ratio (CR) can be calculated as $ CR = \frac{k}{2n} $. 
In this paper, the CR takes values from the set $CR\in\{1/96,2/96,3/96,4/96,5/96,6/96\}$.

\subsubsection{Training Settings}

All experiments are conducted on a single NVIDIA RTX A6000 GPU.
We use a batch size of 128 and adopt the Adam optimizer for training.
%
Unless otherwise stated, the channel SNR for the legitimate user and the eavesdropper is identical, with values of $\rm SNR_{\rm leg}=\rm SNR_{\rm eve}\in\{0,5,10,15,20\}dB$.
The hyperparameters $\lambda_1 $, $\lambda_2$, and $\lambda_3$ are set to 0.01, 1.0, and 1.5, respectively. 
The PAC $a$ takes two values, 0.40 and 0.49, selected based on the SEP analysis in Section IV-C. These values correspond to different security levels. Specifically, $a = 0.49$ provides higher security, while $a = 0.40$ represents a moderate security level.
%
When the private dataset is FFHQ-64, the hyperparameter $N$ in the semantic decoder is set to 2. 
When the private dataset is MNIST, $N$ is set to 1. 
The learning rate for the first and second training stages is set to 
$2 \times 10^{-4}$, with both stages lasting for 100 epochs. 
For the third training stage, the learning rate is set to $5 \times 10^{-5}$, with training lasting for 200 epochs.

\subsubsection{Performance Metric}


In this paper, we use PSNR as a metric to evaluate the reconstruction performance of both the legitimate user and the eavesdropper. 
The higher the PSNR of the legitimate user, the better the system's task performance. 
Conversely, the higher the PSNR of the eavesdropper, the lower the system's security.

\subsubsection{The Effect of Modulation Order}

To facilitate the derivation of SEPs and the implementation of simulations, we set the modulation order of both the outer and inner constellations to 4-QAM.
However, it is important to note that the proposed framework is general and can be easily extended to support higher-order modulation schemes.
When employing higher-order modulation, the number of output neurons in the FC layer of each QAM modulator, as well as the number of input channels in the first convolutional layer of the semantic decoder, should be increased accordingly.
Since the modulation order only affects a limited number of layers in the proposed system, increasing it introduces only a small overhead in training and inference time.
As shown in our previous work \cite{bo2024jcm}, increasing the modulation order from 4-QAM to 16-QAM results in only a 14\% increase in training and inference times. 
Regarding constellation map design, although higher-order modulation complicates the derivation of SEPs, the added complexity is manageable and does not affect the system's training or inference efficiency.

\subsubsection{Choice of Neural Network Architecture}

We adopt a convolutional neural network (CNN) architecture as the main component of both the semantic encoder and decoder, as CNNs are well suited for image-based tasks and are widely used in existing secure SemCom studies \cite{tung2023deep,meng2025secure}.
Although more complex architectures have recently gained popularity in SemCom, we do not adopt them here due to the wiretap channel scenario considered in this work, where the eavesdropper is assumed to have stolen the architecture and parameters of the legitimate user's semantic decoder.
In this case, using a more powerful architecture would improve the performance of both users and increase the computational complexity of the system, neither of which contributes to system security.
Since our focus is on enhancing system security rather than solely maximizing task performance, the CNN architecture is sufficient for our purposes.
Moreover, it is worth noting that the proposed method can be readily applied to other neural network architectures.

\subsubsection{The Benchmarks}


We use two benchmarks to demonstrate the performance and security of our proposed system.
First, we adopt the ESCS method proposed by Luo \textit{et al.} \cite{luo2023encrypted} as the initial benchmark for performance comparison, as it shares a similar setup with our proposed system and currently represents the state-of-the-art under such conditions. 
Both methods assume that the transmitter and the legitimate user share certain prior knowledge in advance, and neither relies on an SNR gap between the legitimate user and the eavesdropper.
ESCS is configured in encryption mode, and we train it using the adversarial training approach.
All other settings are consistent with those used in our method.

For the second benchmark, we use our previous work \cite{chen2024nearly} to highlight the advantages brought by our proposed approach. 
Our previous work relies on a hyperparameter, the minimum SEP (MSEP) of the eavesdropper.
The larger the MSEP, the higher the system security. 
We select MSEP = 74\% and MSEP = 73\% as benchmark cases. 
Since our previous work depends on an SNR gap between the legitimate user and the eavesdropper, 
we align the channel conditions for comparison with our previous work by setting $\rm SNR_{\rm leg}=20dB$ and $\rm SNR_{\rm eve}\in\{-15,-10,-5,0,5\}dB$. 
Additionally, to ensure consistency with our previous work, we use the CIFAR-10 dataset for training and testing.

\subsection{Training and Inference Times of the Proposed System}

In this subsection, we present the average training time per image for each epoch, as well as the total training time per epoch across different training stages of the proposed system, all measured on a single NVIDIA RTX A6000 GPU. Additionally, we report the corresponding results for the inference stage.
The CIFAR-100 dataset is used for the source images, and the MNIST dataset is used for the private images.
The PAC is set to 0.49, the compression ratio is set to 1/24, and $\rm SNR_{\rm leg}=\rm SNR_{\rm eve}=10dB$.
The experimental results are shown in Table~\ref{tab1-1-24}. It can be observed that training stage 3 requires more time than stage 1 and stage 2, as it involves training the entire model.
The proposed system achieves an inference time of 0.673 ms per image, which corresponds to a processing throughput of approximately 1,486 image frames per second. This demonstrates the strong real-time communication capability of the system.
Overall, the experimental results confirm the scalability of the proposed system, indicating that it can be easily extended to more complex image-based tasks.


\begin{table}[h]
\caption{Training and inference times of the proposed system on a single NVIDIA RTX A6000 GPU. The PAC value is set to 0.49, the private dataset is MNIST, the compression ratio is 1/24, and $\rm SNR_{\rm leg}=\rm SNR_{\rm eve}=10dB$.}
\centering
\begin{tabular}{c|c|c}
\hline
Stage & \begin{tabular}[c]{@{}c@{}}Time per\\ image (ms)\end{tabular}& \begin{tabular}[c]{@{}c@{}}Total time\\ per epoch (s)\end{tabular} \\ \hline
Training (Stage 1)  & 0.804 ms  & 40.2 s  \\ \hline
Training (Stage 2)  & 0.555 ms  & 27.7 s  \\ \hline
Training (Stage 3)  & 1.168 ms  & 58.4 s  \\ \hline
Inference           & 0.673 ms  & 6.7 s  \\ \hline
\end{tabular}
\label{tab1-1-24}
\vskip -0.1in
\end{table}

\subsection{Performance Comparison of Different Approaches}

In this subsection, we compare the performance of our proposed system against existing methods under various channel SNRs and compression ratios.
The comparison focuses on both the security and reconstruction performance of the system.
The security of the system is evaluated based on the PSNR performance of the eavesdropper, where a lower PSNR indicates higher system security.
The task performance of the system is evaluated based on the PSNR performance of the legitimate user, with a higher PSNR indicating better reconstruction quality.
In the first two experiments, the channel SNR is identical for both the legitimate user and the eavesdropper, and we use the CIFAR-100 dataset for source image transmission.
In the third experiment, the channel SNR differs between the legitimate user and the eavesdropper, and we use the CIFAR-10 dataset for source image transmission.

\begin{figure}[t]
\begin{center}
\centerline{\includegraphics[width=0.95\linewidth]{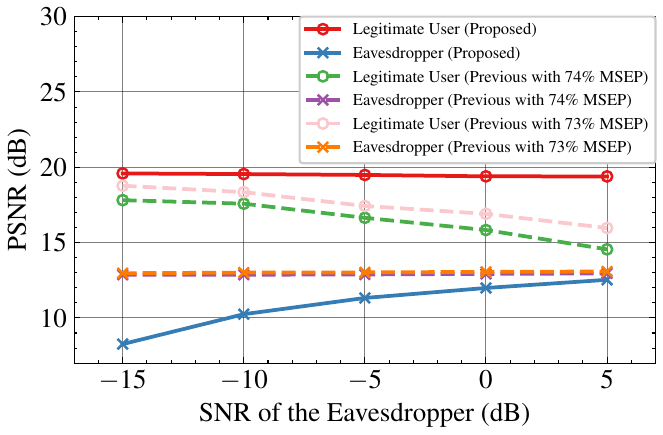}}
\caption{The PSNR performance of our proposed method and our previous work for both users under different $\rm SNR_{\rm eve}$. The PAC value is set to 0.49, the private dataset is FFHQ-64, the compression ratio is 1/24, and $\rm SNR_{\rm leg}=20dB$.}
\label{fig.previous_benchmark}
\end{center}
\vskip -0.3in
\end{figure}

\begin{figure}[htbp]
\centering
\subfigure[$a=0.49$]{
\begin{minipage}[t]{1\linewidth}
\centering
\includegraphics[width=0.95\linewidth]{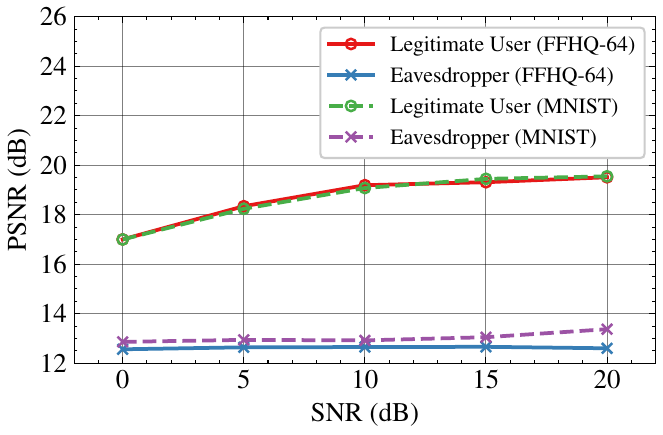}
\end{minipage}
}
\subfigure[$a=0.40$]{
\begin{minipage}[t]{1\linewidth}
\centering
\includegraphics[width=0.95\linewidth]{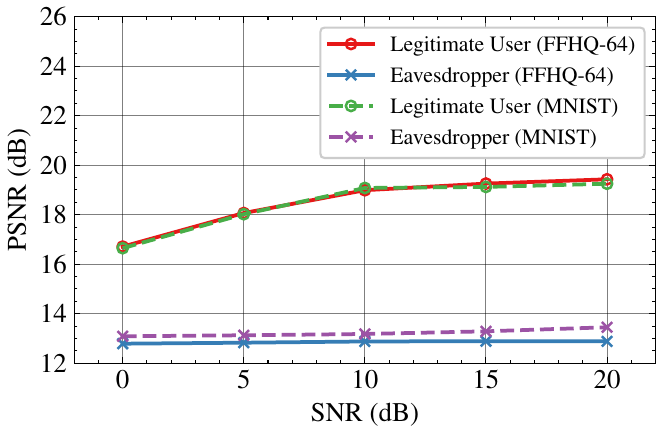}
\end{minipage}
}
\caption{The PSNR performance of our proposed method for both users under different channel SNRs and private datasets, with PAC values of 0.49 and 0.40. The private datasets used are FFHQ-64 and MNIST, the compression ratio is 1/24, and $\rm SNR_{\rm leg}=\rm SNR_{\rm eve}$.
}
\label{fig.plot3}
\vskip -0.2in
\end{figure}

\subsubsection{Different Channel SNRs}


Fig.~\ref{fig.plot1} shows the PSNR performance of both the eavesdropper and the legitimate user in our proposed system, as well as in ESCS, under different channel SNRs.
The compression ratio is set to 1/24, with the private dataset being FFHQ-64, and the PAC values are set to 0.49 and 0.40, respectively. 
From Fig.~\ref{fig.plot1}, it is clear that our proposed system significantly outperforms the benchmark in terms of reconstruction performance, although its security is slightly lower than that of ESCS.
Furthermore, as the channel SNR increases, the PSNR performance of the legitimate user in our proposed system improves more rapidly than that of ESCS, while the PSNR performance of the eavesdropper remains nearly unchanged, similar to the behavior observed in ESCS.
For example, with a PAC value of 0.49, the PSNR performance of the eavesdropper in our proposed system is approximately 0.3 dB higher than that of ESCS, while the PSNR performance of the legitimate user is about 1.0-1.9 dB higher.
This clearly demonstrates that our proposed system significantly outperforms the benchmark in terms of task performance across various channel SNRs, while maintaining comparable security levels.

\subsubsection{Different Compression Ratios}


Fig.~\ref{fig.plot2} shows the PSNR performance of both the eavesdropper and the legitimate user in our proposed system, as well as in ESCS, under different compression ratios. 
The channel SNR is set to 10 dB, the private dataset is FFHQ-64, and the PAC values are set to 0.49 and 0.40, respectively. 
From Fig.~\ref{fig.plot2}, it is evident that, across all compression ratios, our proposed system significantly outperforms ESCS in terms of task performance, while the PSNR performance of the eavesdropper in our proposed system is slightly higher than in ESCS.
Additionally, as the compression ratio increases, the PSNR performance of the legitimate user in our proposed system improves more rapidly compared to ESCS, while the PSNR performance of the eavesdropper improves gradually and only marginally compared to ESCS.
For example, with PAC = 0.49, across all compression ratios, the PSNR performance of the legitimate user in our proposed system is about 1.7-2.2 dB higher than in ESCS, while the PSNR performance of the eavesdropper is only 0.2-0.9 dB higher. 
This minor increase in the PSNR performance of the eavesdropper does not lead to a significant visual improvement, as Eve's PSNR remains very low, making it nearly impossible for the eavesdropper to extract valid information.
These results clearly demonstrate that our proposed system significantly outperforms the benchmark in terms of task performance across various compression ratios, while maintaining comparable security levels.

\begin{figure}[t]
\centering
\subfigure[$a=0.49$]{
\begin{minipage}[t]{1\linewidth}
\centering
\includegraphics[width=0.95\linewidth]{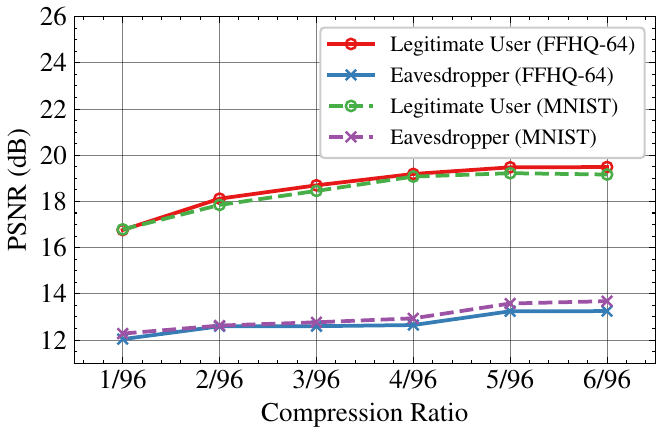}
\end{minipage}
}
\subfigure[$a=0.40$]{
\begin{minipage}[t]{1\linewidth}
\centering
\includegraphics[width=0.95\linewidth]{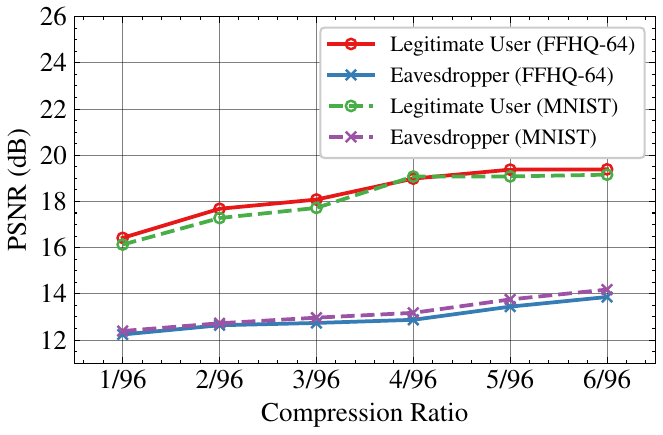}
\end{minipage}
}
\caption{The PSNR performance of our proposed method for both users under different compression ratios and private datasets, with PAC values of 0.49 and 0.40. The private datasets used are FFHQ-64 and MNIST, the channel SNR is 10 dB, and $\rm SNR_{\rm leg}=\rm SNR_{\rm eve}$.
}
\label{fig.plot4}
\vskip -0.2in
\end{figure}

\begin{figure}[t]
\centering
\subfigure[The private dataset is FFHQ-64.]{
\begin{minipage}[t]{1\linewidth}
\centering
\includegraphics[width=0.95\linewidth]{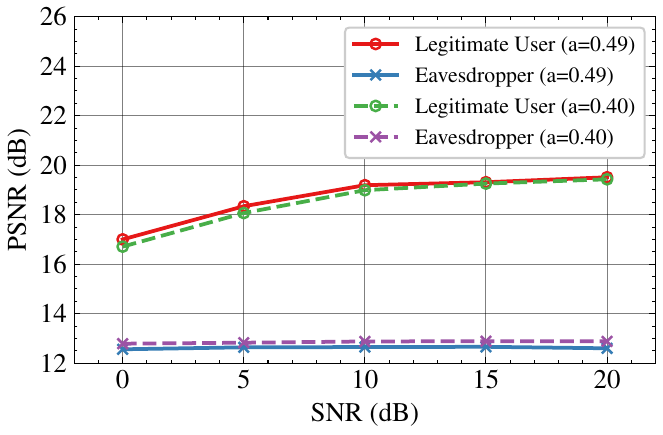}
\end{minipage}
}
\subfigure[The private dataset is MNIST.]{
\begin{minipage}[t]{1\linewidth}
\centering
\includegraphics[width=0.95\linewidth]{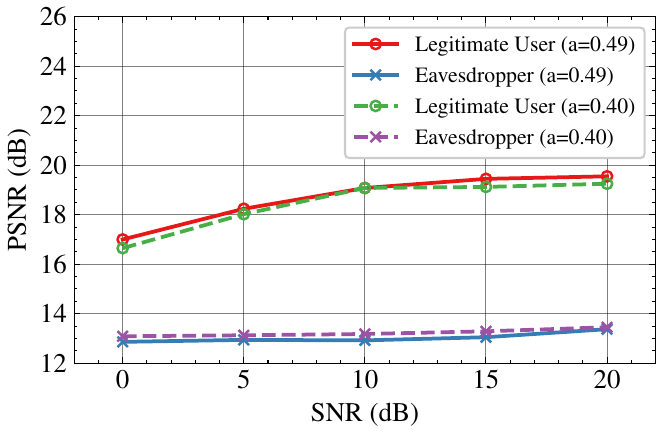}
\end{minipage}
}
\caption{The PSNR performance of our proposed method for both users under different channel SNRs and PAC values, using the FFHQ-64 and MNIST private datasets. The PAC values are 0.49 and 0.40, the compression ratio is 1/24, and $\rm SNR_{\rm leg}=\rm SNR_{\rm eve}$.
}
\label{fig.plot5}
\vskip -0.2in
\end{figure}

\subsubsection{Different Channel SNR Gaps between Bob and Eve}

Fig.~\ref{fig.previous_benchmark} presents the PSNR performance of both the eavesdropper and the legitimate user in our proposed system, compared to our previous work \cite{chen2024nearly}, under different channel SNR gaps between the users. 
We set $\rm SNR_{\rm leg}=20dB$ and $\rm SNR_{\rm eve}\in\{-15,-10,-5,0,5\}dB$. 
The source dataset is CIFAR-10, the private dataset is FFHQ-64, the PAC value is 0.49, and the compression ratio is 1/24.
From Fig.~\ref{fig.previous_benchmark}, it is clear that as the channel SNR of the eavesdropper improves, the PSNR performance of the legitimate user in our proposed system remains nearly constant, while it gradually decreases in our previous work. 
This indicates that our proposed system provides a significant advantage in task performance.
Additionally, although the PSNR performance of the eavesdropper increases as the channel SNR of the eavesdropper improves in our proposed system, it remains consistently lower than in our previous work across all channel conditions.
This indicates that the security of our proposed system is always superior to that of our previous work, regardless of the SNR gap between the users.
Overall, compared to our previous work, our proposed system offers clear benefits in both system security and task performance, with a particularly notable improvement in the latter, highlighting the effectiveness of our proposed jamming approach.

\subsection{Performance Comparison of Different Hyperparameters}


In this subsection, we compare the effects of different private datasets and PAC values on the security and task performance of our proposed system under different channel SNRs and compression ratios.

\subsubsection{Different Private Datasets}


Figs.~\ref{fig.plot3} and \ref{fig.plot4} show the PSNR performance of both the legitimate user and the eavesdropper in our proposed system using two different private datasets, FFHQ-64 and MNIST, under different channel SNRs and compression ratios. 
From Fig.~\ref{fig.plot3}, it is evident that the PSNR performance of the legitimate user remains similar whether FFHQ-64 or MNIST is used as the private dataset.
However, for the eavesdropper, the PSNR performance with FFHQ-64 as the private dataset is generally lower than with MNIST, with a difference of approximately 0.3-0.8 dB.
In Fig.~\ref{fig.plot4}, under different compression ratios, 
the PSNR performance of the legitimate user with FFHQ-64 as the private dataset is about 0.1-0.4 dB higher than with MNIST. 
For the eavesdropper, the PSNR performance with FFHQ-64 as the private dataset is generally lower than with MNIST by around 0.1-0.4 dB.
These results indicate that a more complex private dataset (such as FFHQ-64) provides stronger interference, leading to a lower PSNR for the eavesdropper. In contrast, a less complex dataset (such as MNIST) results in weaker interference, 
which leads to a smaller performance gap between the legitimate user and the eavesdropper.
Therefore, this demonstrates that using a high-complexity private dataset enhances the security of our proposed system.

\subsubsection{Different Power Allocation Coefficients}


\begin{figure}[t]
\centering
\subfigure[The private dataset is FFHQ-64.]{
\begin{minipage}[t]{1\linewidth}
\centering
\includegraphics[width=0.95\linewidth]{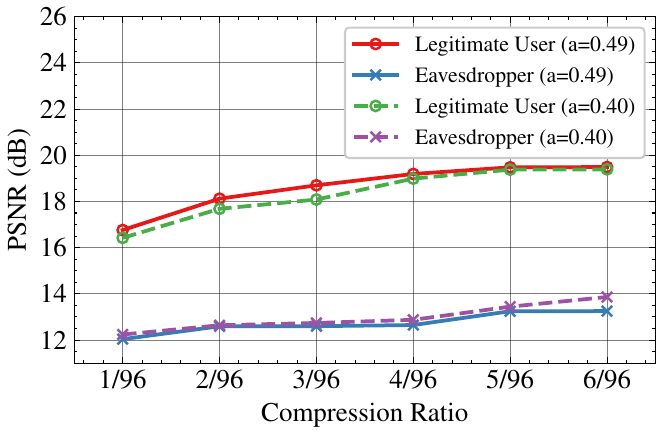}
\end{minipage}
}
\subfigure[The private dataset is MNIST.]{
\begin{minipage}[t]{1\linewidth}
\centering
\includegraphics[width=0.95\linewidth]{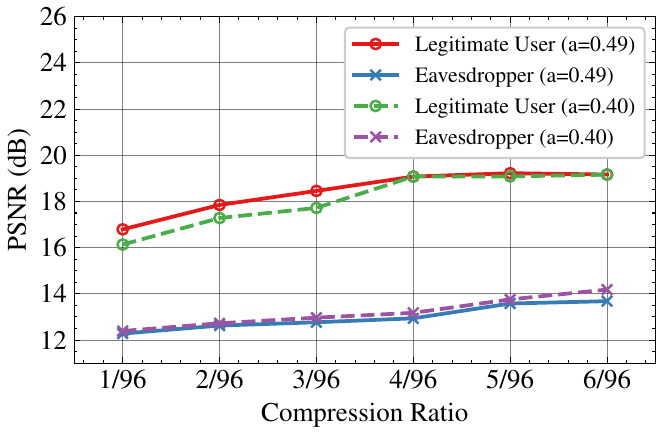}
\end{minipage}
}
\caption{The PSNR performance of our proposed method for both users under different compression ratios and PAC values, using the FFHQ-64 and MNIST private datasets. The PAC values are 0.49 and 0.40, the channel SNR is 10 dB, and $\rm SNR_{\rm leg}=\rm SNR_{\rm eve}$.
}
\label{fig.plot6}
\vskip -0.2in
\end{figure}

Figs.~\ref{fig.plot5} and \ref{fig.plot6} show the PSNR performance of both the legitimate user and the eavesdropper in our proposed system, using two different PAC values (0.49 and 0.40) under various channel SNRs and compression ratios. 
From Fig.~\ref{fig.plot5}, it can be observed that, under different channel SNRs, the legitimate user's PSNR with a PAC value of 0.49 is slightly higher than with a PAC value of 0.40. Additionally, the eavesdropper's PSNR with a PAC value of 0.49 is also slightly lower than with a PAC value of 0.40. 
In Fig.~\ref{fig.plot6}, similar conclusions are drawn for different compression ratios. These results suggest that a higher PAC value provides better security and task performance for our proposed system, 
while reducing the PAC to 0.40 leads to a decline in both system security and task performance. 
This validates our theoretical analysis in Section IV and demonstrates that adjusting the PAC values allows for flexible and explicit control over system security.


\begin{figure*}[htbp]
\centering
\subfigure[Source image]{
\begin{minipage}[t]{0.19\linewidth}
\centering
\includegraphics[width=0.8in]{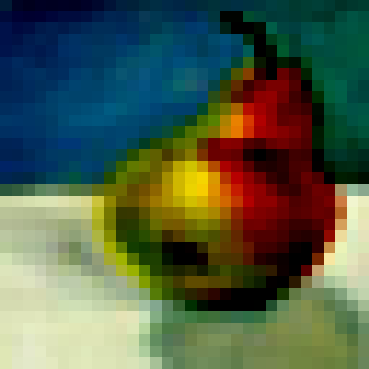}
\end{minipage}%
}%
\subfigure[Bob (Proposed)]{
\begin{minipage}[t]{0.19\linewidth}
\centering
\includegraphics[width=0.8in]{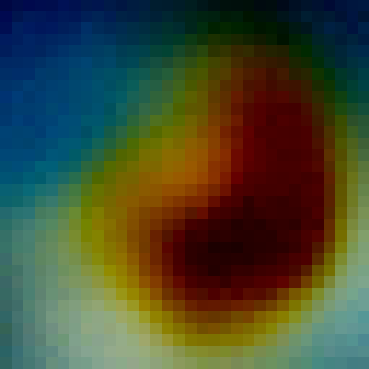}
\end{minipage}%
}%
\subfigure[Eve (Proposed)]{
\begin{minipage}[t]{0.19\linewidth}
\centering
\includegraphics[width=0.8in]{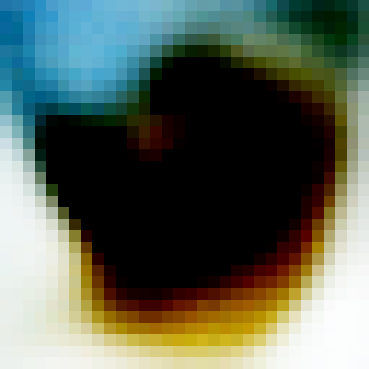}
\end{minipage}
}%
\subfigure[Bob (ESCS)]{
\begin{minipage}[t]{0.19\linewidth}
\centering
\includegraphics[width=0.8in]{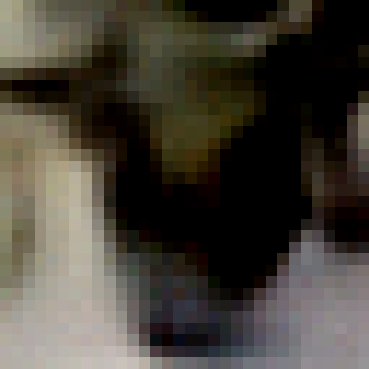}
\end{minipage}
}%
\subfigure[Eve (ESCS)]{
\begin{minipage}[t]{0.19\linewidth}
\centering
\includegraphics[width=0.8in]{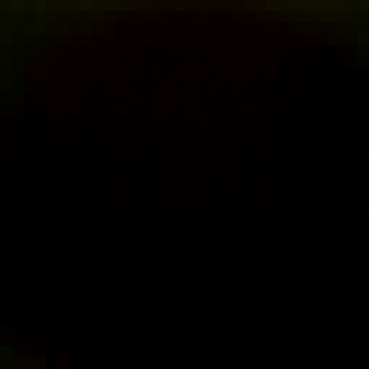}
\end{minipage}
}%
\centering
\caption{Visual analysis of the recovered images for both users of the proposed method and ESCS, with a channel SNR of 10 dB. 
The PAC is 0.49, the private dataset is FFHQ-64, and the compression ratio is 1/24. 
In addition, $\rm SNR_{\rm leg}=\rm SNR_{\rm eve}$. 
}
\label{fig.visual}
\vskip -0.2in
\end{figure*}

\subsection{Visual Comparison}


We also conduct a visual analysis of the recovered images of both the legitimate user and the eavesdropper in our proposed system and ESCS. 
The experiment is conducted with a channel SNR of 10 dB, PAC set to 0.49, the private dataset being FFHQ-64, and a compression ratio of 1/24.
As shown in Fig.~\ref{fig.visual}, (a) is the source image, (b) and (c) represent the recovered images of the legitimate user and eavesdropper in our proposed system, while (d) and (e) represent the recovered images of the legitimate user and eavesdropper in ESCS.
First, by comparing (a), (c), and (e), we observe that the recovered images of the eavesdropper in both our proposed system and ESCS show a significant difference from the source image, indicating that both eavesdroppers are nearly unable to extract valid information. This suggests that the security levels of both systems are similar. 
Next, comparing (a), (b), and (d), we find that the recovered image of the legitimate user in our proposed system is visually closer to the source image compared to the eavesdropper's.
These results demonstrate that our proposed system not only achieves a security level comparable to ESCS but also delivers superior visual performance.

\subsection{Ablation Study on nHSIC Term}

\begin{figure}[t]
\begin{center}
\centerline{\includegraphics[width=0.95\linewidth]{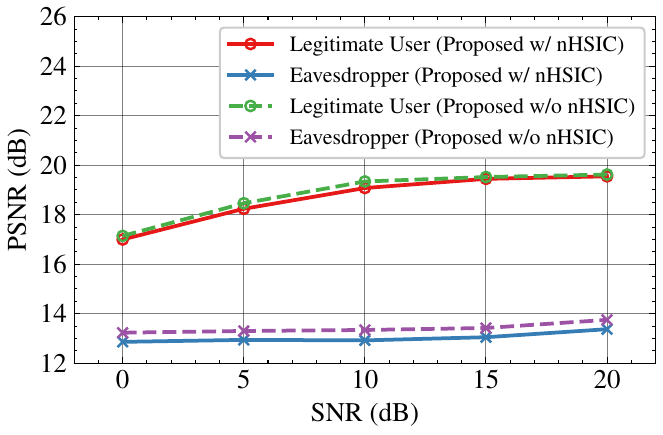}}
\caption{The PSNR performance of the proposed system with and without the nHSIC term for both users under different channel SNRs.
The PAC value is set to 0.49, the private dataset is MNIST, the compression ratio is 1/24, and $\rm SNR_{\rm leg}=\rm SNR_{\rm eve}$.}
\label{fig.nHSIC}
\end{center}
\vskip -0.3in
\end{figure}

In this paper, we incorporate an nHSIC term into the loss function. This term is introduced to minimize the similarity between the semantic information $\textbf{Y}_1$ and the jamming signal $\textbf{Y}_2$, thereby reducing information leakage caused by semantic similarity between the source image and the private image. This design further enhances the effectiveness of the jamming signal in improving system security.
In this subsection, we conduct an ablation study on the nHSIC term to evaluate its specific contribution. We compare the PSNR performance of both the legitimate user and the eavesdropper under the proposed system, with (w/) and without (w/o) the nHSIC term, across various channel SNRs.
The source dataset is CIFAR-100, and the private dataset is MNIST. The PAC is set to 0.49, and the compression ratio is fixed at 1/24. We have $\mathrm{SNR}_{\mathrm{leg}} = \mathrm{SNR}_{\mathrm{eve}} \in \{0, 5, 10, 15, 20\}~\mathrm{dB}$. The experimental results are shown in Fig.~\ref{fig.nHSIC}. 
From Fig.~\ref{fig.nHSIC}, we observe that, in terms of PSNR performance for the legitimate user, the system without the nHSIC term achieves slightly better results than the system with the nHSIC term across all channel SNRs.
However, in terms of system security, the system with the nHSIC term noticeably outperforms the system without the nHSIC term.
The experimental results indicate that incorporating the nHSIC term into the loss function effectively reduces the semantic similarity between the semantic information and the jamming signal.
This reduction in semantic similarity leads to decreased PSNR performance for both the legitimate user and the eavesdropper, as less source data-related semantic information is preserved in the transmitted signal.
Notably, the PSNR degradation for the eavesdropper is more significant, demonstrating the effectiveness of the nHSIC term in enhancing system security.


\subsection{Complexity Analysis of the Proposed System with and without Security Enhancement}

The proposed method enhances the security of SemCom systems. However, the introduction of several modules for security enhancement inevitably increases the overall complexity of the system. In this subsection, we analyze the complexity of the proposed system with and without security enhancement, considering both computational complexity and the number of parameters. Computational complexity is measured using floating-point operations (FLOPs), which indicate the number of arithmetic operations required by the model.
The proposed system without security enhancement refers to the removal of the inner semantic encoders at both the transmitter and the receiver, along with their corresponding 4-QAM modulators. The source dataset is CIFAR-100, and the private dataset is MNIST. The compression ratio is set to 1/24. The experimental results are shown in Table~\ref{tab:flops-param}.
We observe that the proposed system with security enhancement exhibits higher complexity due to the inclusion of dedicated modules for enhancing security.
Specifically, the system without security enhancement requires 55.6\% of the FLOPs and 65.1\% of the parameter count compared to the system with security enhancement. 
Despite this increase in complexity, the proposed system with security enhancement still maintains high computational efficiency, with an inference time of just 0.673 ms per image (Table~\ref{tab1-1-24}). 
More importantly, this additional complexity can lead to significantly improved system security. 

\begin{table}[h]
\caption{FLOPs and parameter count of the proposed system with and without security enhancement. The source dataset is CIFAR-100, and the private dataset is MNIST. The compression ratio is 1/24.}
\centering
\begin{tabular}{c|c|c}
\hline
Model & FLOPs (G) & \begin{tabular}[c]{@{}c@{}}Parameter\\Count (M)\end{tabular} \\ \hline
Proposed w/ security enhancement    & 906.58 G  & 99.2 M   \\ \hline
Proposed w/o security enhancement & 504.02 G  & 64.6 M   \\ \hline
\end{tabular}
\label{tab:flops-param}
\vskip -0.1in
\end{table}

\section{Conclusion}

In this paper, we proposed a secure SemCom approach for image transmission that remains effective even when the legitimate and eavesdropping channels exhibit similar conditions. 
By leveraging shared prior knowledge for superposition coding, our method generates two 4-QAM constellation sequences: (1) an outer sequence carrying semantic information and (2) an inner sequence derived from a private image in a shared private database between Alice and Bob, which acts as a jamming signal. 
We further introduced a power allocation coefficient (PAC) to dynamically control interference based on the derived SEPs of both the legitimate user and the eavesdropper, thereby enabling explicitly controllable enhancement of system security.
During training, mutual information (MI) minimization strengthens the jamming effect, ensuring that only the legitimate receiver, who possesses the private image, can successfully decode the transmitted data. 
Numerical results validated the effectiveness of our approach, demonstrating a security level comparable to that of the ESCS method and achieving a task performance gain of over 1 dB, thereby underscoring its practical superiority. 
%
%
Moreover, the proposed shared knowledge-based jamming approach via superposition coding can also be easily extended to other SemCom tasks \cite{ng2022ninjadesc,lo2023collaborative} and applied to enhance transmission security across various modalities, simply by correspondingly adapting the architectures of the semantic encoder and decoder.
%
Although the proposed method was developed and evaluated for digital SemCom systems in this paper, the shared knowledge-based jamming approach can also be extended to analog DeepJSCC schemes.
In such scenarios, the configuration of PAC needs to be guided by specific optimization algorithms or reinforcement learning (RL) agents, rather than relying on SEP-based derivation and analysis.
Future work may explore adaptive PAC optimization under dynamic channel conditions, or the extension of the proposed framework to broader application scenarios \cite{zhang2023energy,zhang2024generative,zhang2025covert}.

\bibliographystyle{IEEEtran}

\bibliography{myref}

@inproceedings{ng2022ninjadesc,
  author       = {Tony Ng and
                  Hyo Jin Kim and
                  Vincent T. Lee and
                  Daniel DeTone and
                  Tsun{-}Yi Yang and
                  Tianwei Shen and
                  Eddy Ilg and
                  Vassileios Balntas and
                  Krystian Mikolajczyk and
                  Chris Sweeney},
  title        = {NinjaDesc: Content-Concealing Visual Descriptors via Adversarial Learning},
  booktitle    = {Proc. {IEEE/CVF} {CVPR}, New Orleans, LA, USA, Jun. 2022},
  pages        = {12787--12797},
}

@article{lo2023collaborative,
   author       = {Wing Fei Lo and
                  Nitish Mital and
                  Haotian Wu and
                  Deniz G{\"{u}}nd{\"{u}}z},
  title        = {Collaborative Semantic Communication for Edge Inference},
  journal      = {{IEEE} Wirel. Commun. Lett.},
  volume       = {12},
  number       = {7},
  pages        = {1125--1129},
  year         = {Mar. 2023},
}

@article{meng2025secure,
  title={Secure Semantic Communication With Homomorphic Encryption}, 
  author={Rui Meng and Dayu Fan and Haixiao Gao and Yifan Yuan and Bizhu Wang and Xiaodong Xu and Mengying Sun and Chen Dong and Xiaofeng Tao and Ping Zhang and Dusit Niyato},
  journal={arXiv:2501.10182v1 [cs.CR]},
  year={Jan. 2025}
}

@article{bo2024jcm,
  author       = {Yufei Bo and
                  Yiheng Duan and
                  Shuo Shao and
                  Meixia Tao},
  title        = {Joint Coding-Modulation for Digital Semantic Communications via Variational
                  Autoencoder},
  journal      = {{IEEE} Trans. Commun.},
  volume       = {72},
  number       = {9},
  pages        = {5626--5640},
  year         = {Apr. 2024}
}

@article{chen2024nearly,
  author       = {Weixuan Chen and
                  Shuo Shao and
                  Qianqian Yang and
                  Zhaoyang Zhang and
                  Ping Zhang},
  title        = {A Superposition Code-Based Semantic Communication Approach with Quantifiable and Controllable Security},
  journal      = {{IEEE} Trans. Mob. Comput. Early Access},
  volume       = {},
  number       = {},
  pages        = {1--18},
  year         = {Sep. 2025},
}

@article{zhang2025covert,
  title={Covert Prompt Transmission for Secure Large Language Model Services},
  author={Ruichen Zhang and Yinqiu Liu and Shunpu Tang and Jiacheng Wang and Dusit Niyato and Geng Sun and Yonghui Li and Sumei Sun},
  journal={arXiv:2504.21311v1 [cs.NI]},
  year={Apr. 2025}
}

@article{zhang2024generative,
  author       = {Ruichen Zhang and
                  Hongyang Du and
                  Yinqiu Liu and
                  Dusit Niyato and
                  Jiawen Kang and
                  Zehui Xiong and
                  Abbas Jamalipour and
                  Dong In Kim},
  title        = {Generative {AI} Agents With Large Language Model for Satellite Networks
                  via a Mixture of Experts Transmission},
  journal      = {{IEEE} J. Sel. Areas Commun.},
  volume       = {42},
  number       = {12},
  pages        = {3581--3596},
  year = {Sep. 2024}
}

@article{zhang2023energy,
  author       = {Ruichen Zhang and
                  Ke Xiong and
                  Yang Lu and
                  Pingyi Fan and
                  Derrick Wing Kwan Ng and
                  Khaled B. Letaief},
  title        = {Energy Efficiency Maximization in RIS-Assisted {SWIPT} Networks With {RSMA:} {A} PPO-Based Approach},
  journal      = {{IEEE} J. Sel. Areas Commun.},
  volume       = {41},
  number       = {5},
  pages        = {1413--1430},
  year={Jan. 2023}
}

@article{huo2017jamming,
  author       = {Yan Huo and
                  Yuqi Tian and
                  Liran Ma and
                  Xiuzhen Cheng and
                  Tao Jing},
  title        = {Jamming Strategies for Physical Layer Security},
  journal      = {{IEEE} Wirel. Commun.},
  volume       = {25},
  number       = {1},
  pages        = {148--153},
  year         = {Oct. 2017},
}

@article{zhang2024semanticsurvey,
  author       = {Ping Zhang and
                  Wenjun Xu and
                  Yiming Liu and
                  Xiaoqi Qin and
                  Kai Niu and
                  Shuguang Cui and
                  Guangming Shi and
                  Zhijin Qin and
                  Xiaodong Xu and
                  Fengyu Wang and
                  Yue Meng and
                  Chen Dong and
                  Jincheng Dai and
                  Qianqian Yang and
                  Yaping Sun and
                  Dahua Gao and
                  Hui Gao and
                  Shujun Han and
                  Xiaodan Song},
  title        = {Intellicise Wireless Networks From Semantic Communications: {A} Survey,
                  Research Issues, and Challenges},
  journal      = {{IEEE} Commun. Surv. Tutorials},
  volume       = {27},
  number       = {3},
  pages        = {2051--2084},
    year         = {Jul. 2025},
}

@article{bo2024deep,
  author       = {Yufei Bo and
                  Shuo Shao and
                  Meixia Tao},
  title        = {Deep Learning-Based Superposition Coded Modulation for Hierarchical
                  Semantic Communications Over Broadcast Channels},
  journal      = {{IEEE} Trans. Commun.},
  volume       = {73},
  number       = {2},
  pages        = {1186--1200},
  year         = {Aug. 2024},
}

@inproceedings{chen2023deep,
  author       = {Weixuan Chen and
                  Yuhao Chen and
                  Qianqian Yang and
                  Chongwen Huang and
                  Qian Wang and
                  Zhaoyang Zhang},
  title        = {Deep Joint Source-Channel Coding for Wireless Image Transmission with
                  Entropy-Aware Adaptive Rate Control},
  booktitle    = {Proc. {IEEE} {GLOBECOM}, Kuala Lumpur,
                  Malaysia, Dec. 2023},
  pages        = {2239--2244},
}

@article{lecun1998gradient,
  author       = {Yann LeCun and
                  L{\'{e}}on Bottou and
                  Yoshua Bengio and
                  Patrick Haffner},
  title        = {Gradient-based learning applied to document recognition},
  journal      = {Proc. {IEEE}},
  volume       = {86},
  number       = {11},
  pages        = {2278--2324},
  year         = {Nov. 1998},
}

@article{karras2019style,
   author       = {Tero Karras and
                  Samuli Laine and
                  Timo Aila},
  title        = {A Style-Based Generator Architecture for Generative Adversarial Networks},
  journal      = {{IEEE} Trans. Pattern Anal. Mach. Intell.},
  volume       = {43},
  number       = {12},
  pages        = {4217--4228},
  year         = {Mar. 2022},
}

@article{zheng2021information,
  title={An Information Theory-inspired Strategy for Automatic Network Pruning}, 
      author={Xiawu Zheng and Yuexiao Ma and Teng Xi and Gang Zhang and Errui Ding and Yuchao Li and Jie Chen and Yonghong Tian and Rongrong Ji},
        journal      = {arXiv:2108.08532v3 [cs.CV]},
      year={Aug. 2021},
}

@inproceedings{jang2016categorical,
  author       = {Eric Jang and
                  Shixiang Gu and
                  Ben Poole},
  title        = {Categorical Reparameterization with Gumbel-Softmax},
  booktitle    = {Proc. {ICLR},
                  Toulon, France, Apr. 2017},
    pages        = {1--6},

}

@inproceedings{bo2022learning,
  author       = {Yufei Bo and
                  Yiheng Duan and
                  Shuo Shao and
                  Meixia Tao},
  title        = {Learning Based Joint Coding-Modulation for Digital Semantic Communication
                  Systems},
  booktitle    = {Proc. 14th {WCSP}, Nanjing, China, Nov. 2022},
  pages        = {1--6},
}

@inproceedings{qin2023securing,
  author       = {Qi Qin and
                  Yankai Rong and
                  Guoshun Nan and
                  Shaokang Wu and
                  Xuefei Zhang and
                  Qimei Cui and
                  Xiaofeng Tao},
  title        = {Securing Semantic Communications with Physical-Layer Semantic Encryption
                  and Obfuscation},
  booktitle    = {Proc. {IEEE} {ICC}, Rome,
                  Italy, May 2023},
  pages        = {5608--5613},
}

@inproceedings{mu2024semantic,
  author       = {Xidong Mu and
                  Yuanwei Liu},
  title        = {Semantic Communication-Assisted Physical Layer Security Over Fading
                  Wiretap Channels},
  booktitle    = {Proc. {IEEE} {ICC}, Denver,
                  CO, USA, Jun. 2024},
  pages        = {2101--2106},
}

@article{li2024secure,
  author       = {Yongkang Li and
                  Zheng Shi and
                  Han Hu and
                  Yaru Fu and
                  Hong Wang and
                  Hongjiang Lei},
  title        = {Secure Semantic Communications: From Perspective of Physical Layer
                  Security},
  journal      = {{IEEE} Commun. Lett.},
  volume       = {28},
  number       = {10},
  pages        = {2243--2247},
  year         = {Sep. 2024},
}

@inproceedings{chen2023model,
  author       = {Yuhao Chen and
                  Qianqian Yang and
                  Zhiguo Shi and
                  Jiming Chen},
  title        = {The Model Inversion Eavesdropping Attack in Semantic Communication
                  Systems},
  booktitle    = {Proc. {IEEE} {GLOBECOM}, Kuala Lumpur,
                  Malaysia, Dec. 2023},
  pages        = {5171--5177}
}

@article{lin2023blockchain,
  author       = {Yijing Lin and
                  Hongyang Du and
                  Dusit Niyato and
                  Jiangtian Nie and
                  Jiayi Zhang and
                  Yanyu Cheng and
                  Zhaohui Yang},
  title        = {Blockchain-Aided Secure Semantic Communication for AI-Generated Content
                  in Metaverse},
  journal      = {{IEEE} Open J. Comput. Soc.},
  volume       = {4},
  pages        = {72--83},
  year         = {Mar. 2023},
}

@article{he2014providing,
  author       = {Xiang He and
                  Aylin Yener},
  title        = {Providing Secrecy With Structured Codes: Two-User Gaussian Channels},
  journal      = {{IEEE} Trans. Inf. Theory},
  volume       = {60},
  number       = {4},
  pages        = {2121--2138},
  year         = {Jan. 2014},
}

@article{hu2017cooperative,
  author       = {Lin Hu and
                  Hong Wen and
                  Bin Wu and
                  Jie Tang and
                  Fei Pan and
                  Runfa Liao},
  title        = {Cooperative-Jamming-Aided Secrecy Enhancement in Wireless Networks
                  With Passive Eavesdroppers},
  journal      = {{IEEE} Trans. Veh. Technol.},
  volume       = {67},
  number       = {3},
  pages        = {2108--2117},
  year         = {Aug. 2017},
}

@article{sun2022ris,
  author       = {Yifu Sun and
                  Kang An and
                  Yonggang Zhu and
                  Gan Zheng and
                  Kai{-}Kit Wong and
                  Symeon Chatzinotas and
                  Haifan Yin and
                  Pengtao Liu},
  title        = {RIS-Assisted Robust Hybrid Beamforming Against Simultaneous Jamming
                  and Eavesdropping Attacks},
  journal      = {{IEEE} Trans. Wirel. Commun.},
  volume       = {21},
  number       = {11},
  pages        = {9212--9231},
  year         = {May 2022},
}

@article{xu2024coding,
  author       = {Hao Xu and
                  Kai{-}Kit Wong and
                  Yinfei Xu and
                  Giuseppe Caire},
  title        = {Coding-Enhanced Cooperative Jamming for Secret Communication: The
                  {MIMO} Case},
  journal      = {{IEEE} Trans. Commun.},
  volume       = {72},
  number       = {5},
  pages        = {2746--2761},
  year         = {Jan. 2024}
}

@inproceedings{lindner2011better,
  author       = {Richard Lindner and
                  Chris Peikert},
  title        = {Better Key Sizes (and Attacks) for LWE-Based Encryption},
  booktitle    = {Proc. {CT-RSA}, San Francisco, CA, USA, Feb. 2011},
  pages        = {319--339},
}

@article{luo2023encrypted,
  author       = {Xinlai Luo and
                  Zhiyong Chen and
                  Meixia Tao and
                  Feng Yang},
  title        = {Encrypted Semantic Communication Using Adversarial Training for Privacy
                  Preserving},
  journal      = {{IEEE} Commun. Lett.},
  volume       = {27},
  pages        = {1486--1490},
  year         = {Apr. 2023},
}

@article{jiang2022wireless,
  author       = {Peiwen Jiang and
                  Chao{-}Kai Wen and
                  Shi Jin and
                  Geoffrey Ye Li},
  title        = {Wireless Semantic Communications for Video Conferencing},
  journal      = {{IEEE} J. Sel. Areas Commun.},
  volume       = {41},
  pages        = {230--244},
  year         = {Nov. 2022},
}

@inproceedings{zhang2022semantic,
  author       = {Zhenguo Zhang and
                  Qianqian Yang and
                  Shibo He and
                  Zhiguo Shi},
  title        = {Semantic Communication Approach for Multi-Task Image Transmission},
  booktitle    = {Proc. 96th {VTC} Fall, London, United
                  Kingdom, Sep. 2022},
  pages        = {1--2},
}

@article{han2022semanticpre,
  author       = {Tianxiao Han and
                  Qianqian Yang and
                  Zhiguo Shi and
                  Shibo He and
                  Zhaoyang Zhang},
  title        = {Semantic-Preserved Communication System for Highly Efficient Speech
                  Transmission},
  journal      = {{IEEE} J. Sel. Areas Commun.},
  volume       = {41},
  pages        = {245--259},
  year         = {Nov. 2022},
}

@article{weng2023deep,
  author       = {Zhenzi Weng and
                  Zhijin Qin and
                  Xiaoming Tao and
                  Chengkang Pan and
                  Guangyi Liu and
                  Geoffrey Ye Li},
  title        = {Deep Learning Enabled Semantic Communications With Speech Recognition
                  and Synthesis},
  journal      = {{IEEE} Trans. Wirel. Commun.},
  volume       = {22},
  number       = {9},
  pages        = {6227--6240},
  year         = {Feb. 2023},
}

@article{bourtsoulatze2019deep,
  author       = {Eirina Bourtsoulatze and
                  David Burth Kurka and
                  Deniz G{\"{u}}nd{\"{u}}z},
  title        = {Deep Joint Source-Channel Coding for Wireless Image Transmission},
  journal      = {{IEEE} Trans. Cogn. Commun. Netw.},
  volume       = {5},
  number       = {3},
  pages        = {567--579},
  year         = {May 2019},
}

@inproceedings{han2022semantictext,
  author       = {Tianxiao Han and
                  Qianqian Yang and
                  Zhiguo Shi and
                  Shibo He and
                  Zhaoyang Zhang},
  title        = {Semantic-aware Speech to Text Transmission with Redundancy Removal},
  booktitle    = {Proc. {IEEE} {ICC} Workshops, Seoul, Korea, May 2022},
  pages        = {717--722},
}

@inproceedings{erdemir2022privacy,
  author       = {Ecenaz Erdemir and
                  Pier Luigi Dragotti and
                  Deniz G{\"{u}}nd{\"{u}}z},
  title        = {Privacy-Aware Communication over a Wiretap Channel with Generative
                  Networks},
  booktitle    = {Proc. {IEEE} {ICASSP}, Virtual and Singapore, May 2022},
  pages        = {2989--2993},
}

@inproceedings{marchioro2020adversarial,
  author       = {Thomas Marchioro and
                  Nicola Laurenti and
                  Deniz G{\"{u}}nd{\"{u}}z},
  title        = {Adversarial Networks for Secure Wireless Communications},
  booktitle    = {Proc. {IEEE} {ICASSP}, Barcelona, Spain, May 2020},
  pages        = {8748--8752},
}

@inproceedings{tung2023deep,
  author       = {Tze{-}Yang Tung and
                  Deniz G{\"{u}}nd{\"{u}}z},
  title        = {Deep Joint Source-Channel and Encryption Coding: Secure Semantic Communications},
  booktitle    = {Proc. {IEEE} {ICC}, Rome,
                  Italy, May 2023},
  pages        = {5620--5625},
}

@article{tang2025towards,
  title={Towards Secure Semantic Communications in the Presence of Intelligent Eavesdroppers},
  author={Shunpu Tang and Yuhao Chen and Qianqian Yang and Ruichen Zhang and Dusit Niyato and Zhiguo Shi},
  journal={arXiv:2503.23103v1 [cs.IT]},
  year={Mar. 2025}
}

\begin{IEEEbiography}[{\includegraphics[width=1in,height=1.25in,clip,keepaspectratio]{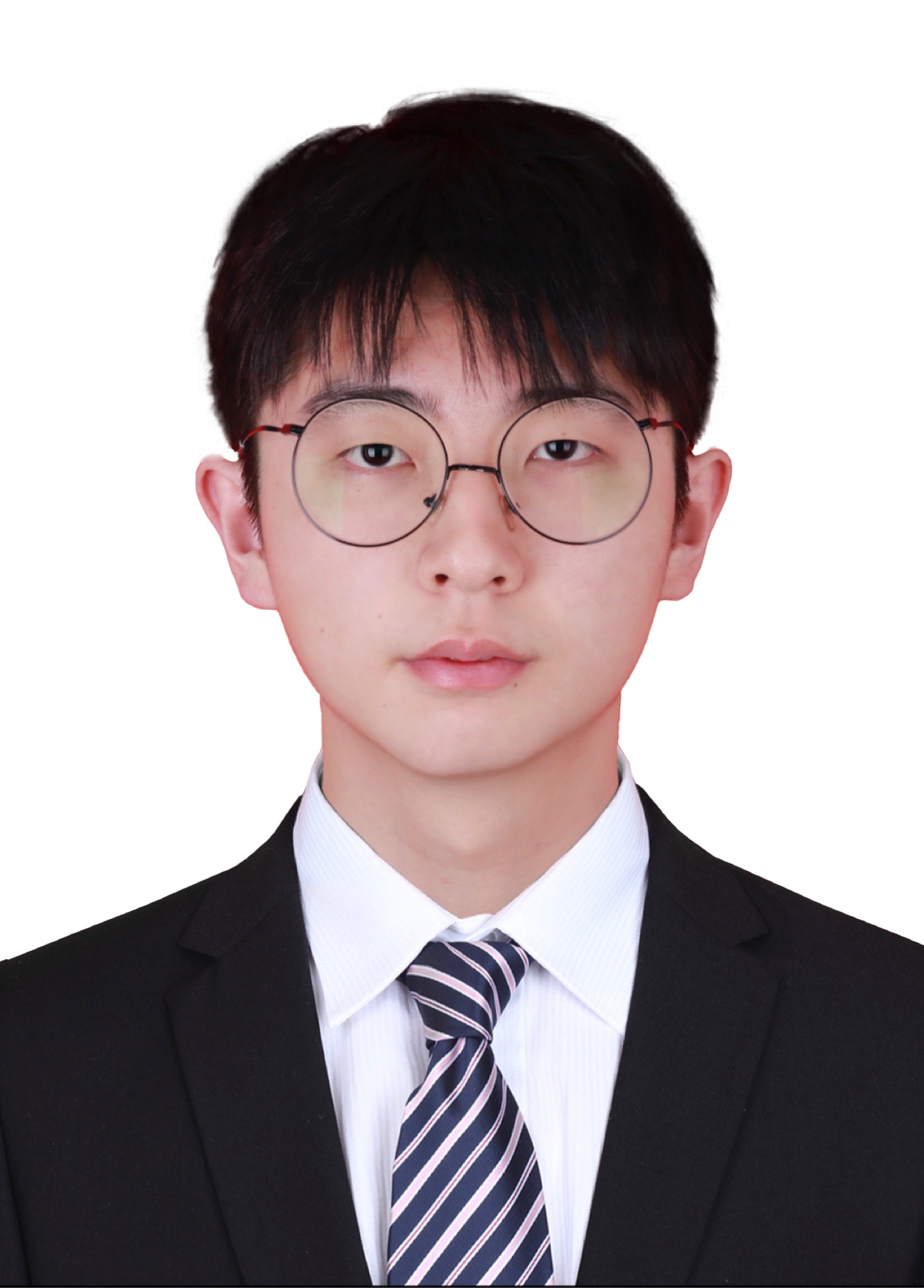}}]{Weixuan Chen}
(Graduate Student Member, IEEE) received the B.E. degree from Sichuan University, Chengdu, China, in 2022.
He is currently working toward the Ph.D. degree with the College of Information Science and Electronic Engineering, Zhejiang University, Hangzhou, China.
His research interests include semantic communications and computer vision.
\end{IEEEbiography}

\begin{IEEEbiography}[{\includegraphics[width=1in,height=1.25in,clip,keepaspectratio]{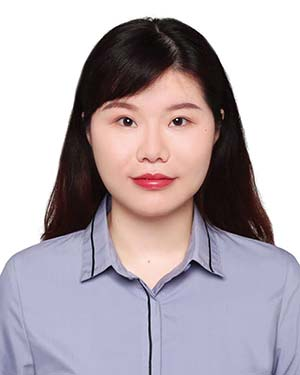}}]{Qianqian Yang}
(Member, IEEE) received the B.Sc. degree in automation from Chongqing University, Chongqing, China, in 2011, the M.S. degree in control engineering from Zhejiang University, Hangzhou, China, in 2014, and the Ph.D. degree in electrical and electronic engineering from Imperial College London, U.K. In 2016, she has held visiting positions with CentraleSupelec and was also with the New York University Tandon School of Engineering from 2017 to 2018. She was a Postdoctoral Research Associate with Imperial College London, and as a Machine Learning Researcher with Sensyne Health Plc. She is currently a Tenure-Tracked Professor with the Department of Information Science and Electronic Engineering, Zhejiang University. Her research interests mainly include wireless communications, information theory, and semantic communications. 
\end{IEEEbiography}

\begin{IEEEbiography}[{\includegraphics[width=1in,height=1.25in,clip,keepaspectratio]{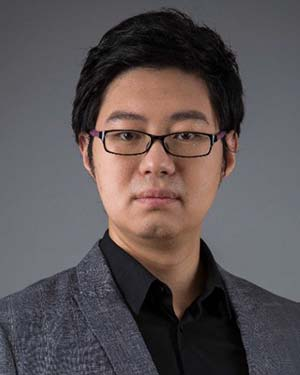}}]{Shuo Shao}
(Member, IEEE) received the B.S. degree in information science from Southeast University, Nanjing, China, in 2011, the M.A.Sc. degree in electrical and computer engineering from McMaster University, Hamilton, ON, Canada, in 2013, and the Ph.D. degree from Texas A\&M University, College Station, TX, USA, in 2017. 
He is currently an Associate Professor with the Department of System Science, University of Shanghai for Science and Technology, and was an Associate Professor with the School of Electronics Information and Electrical Engineering, Shanghai Jiao Tong University, China. His research interests include network information theory, algebraic code, machine learning, and semantic communications.

\end{IEEEbiography}

\begin{IEEEbiography}[{\includegraphics[width=1in,height=1.25in,clip,keepaspectratio]{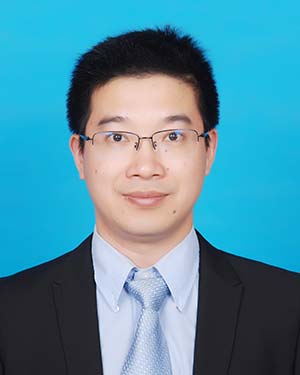}}]{Zhiguo Shi}
(Fellow, IEEE) received the B.S. and Ph.D. degrees in electronic engineering from Zhejiang University, Hangzhou, China, in 2001 and 2006, respectively. Since 2006, he has been a Faculty Member with the College of Information Science and Electronic Engineering, Zhejiang University, where he is currently a Full Professor. From 2011 to 2013, he visited the Broadband Communications Research Group, University of Waterloo, Waterloo, ON, Canada. His research interests include array signal processing, localization, and internet-of-things. Prof. Shi was the recipient of the 2019 IET Communications Premium Award, and coauthored a paper that received the 2021 IEEE Signal Processing Society Young Author Best Paper Award. He was also the recipient of the Best Paper Award from ISAP 2020, IEEE GLOBECOM 2019, IEEE WCNC 2017, IEEE/CIC ICCC 2013, and IEEE WCNC 2013. He was the General Co-Chair of IEEE SAM 2020 and also an Editor for the \textit{IEEE Network} and \textit{IET Communications}. He is currently serving as an Associate Editor for the \textit{IEEE Signal Processing Letters}, \textit{IEEE Transactions on Vehicular Technology}, and \textit{Journal of the Franklin Institute}. He is an elected member of the Sensor Array and Multichannel (SAM) Technical Committee of the IEEE Signal Processing Society.
\end{IEEEbiography}

\begin{IEEEbiography}[{\includegraphics[width=1in,height=1.25in,clip,keepaspectratio]{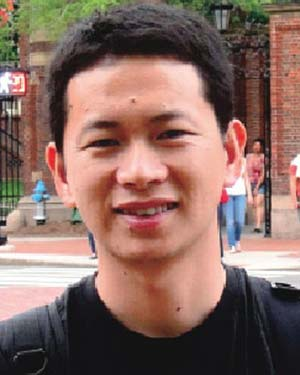}}]{Jiming Chen}
(Fellow, IEEE) received the Ph.D. degree in control science and engineering from Zhejiang University, Hangzhou, China, in 2005.

He is currently a Professor with the Department of Control Science and Engineering, Zhejiang University. His research interests include Internet of Things, networked control, and wireless networks.

Dr. Chen was on the editorial boards of multiple IEEE Transactions, and General Co-Chair for IEEE RTCSA'19, IEEE Datacom'19, and IEEE PST'20.
He was the recipient of the 7th IEEE ComSoc Asia/Pacific Outstanding Paper Award, Japan Society for the Promotion of Science Invitation Fellowship, and IEEE ComSoc AP Outstanding Young Researcher Award. He is also a Distinguished Lecturer of the IEEE Vehicular Technology Society. He is also a Fellow of the CAA.
\end{IEEEbiography}

\begin{IEEEbiography}[{\includegraphics[width=1in,height=1.25in,clip,keepaspectratio]{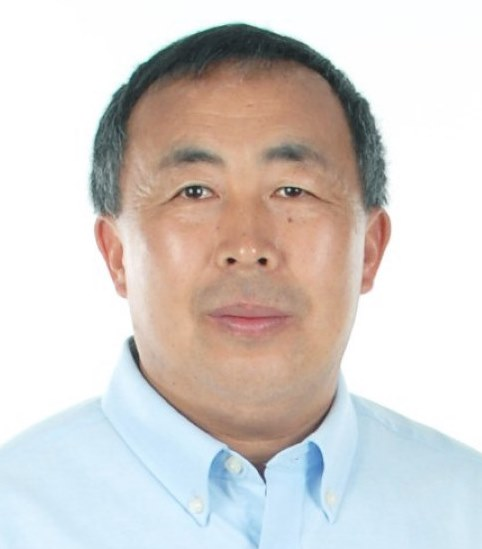}}]{Xuemin (Sherman) Shen}
(Fellow, IEEE) received the Ph.D. degree in electrical engineering from Rutgers University, New Brunswick, NJ, USA, in 1990. He is a University Professor with the Department of Electrical and Computer Engineering, University of Waterloo, Canada. His research focuses on network resource management, wireless network security, Internet of Things, 5G and beyond, and vehicular networks. Dr. Shen is a registered Professional Engineer of Ontario, Canada, an Engineering Institute of Canada Fellow, a Canadian Academy of Engineering Fellow, a Royal Society of Canada Fellow, a Chinese Academy of Engineering Foreign Member, and an International Fellow of the Engineering Academy of Japan.

Dr. Shen received ``West Lake Friendship Award'' from Zhejiang Province in 2023, President's Excellence in Research from the University of Waterloo in 2022, the Canadian Award for Telecommunications Research from the Canadian Society of Information Theory (CSIT) in 2021, the R.A. Fessenden Award in 2019 from IEEE, Canada, Award of Merit from the Federation of Chinese Canadian Professionals (Ontario) in 2019, James Evans Avant Garde Award in 2018 from the IEEE Vehicular Technology Society, Joseph LoCicero Award in 2015 and Education Award in 2017 from the IEEE Communications Society (ComSoc), and Technical Recognition Award from Wireless Communications Technical Committee in 2019 and AHSN Technical Committee in 2013. He has also received the Excellent Graduate Supervision Award in 2006 from the University of Waterloo and the Premier's Research Excellence Award (PREA) in 2003 from the Province of Ontario, Canada.
He serves/served as the General Chair for the 6G Global Conference'23, and ACM Mobihoc'15, Technical Program Committee Chair/Co-Chair for IEEE Globecom'24, 16 and 07, IEEE Infocom'14, IEEE VTC'10 Fall, and the Chair for the IEEE ComSoc Technical Committee on Wireless Communications.
Dr. Shen is the Past President of the IEEE ComSoc, the Vice President for Technical \& Educational Activities, Vice President for Publications, Memberat-Large on the Board of Governors, Chair of the Distinguished Lecturer Selection Committee, and Member of IEEE Fellow Selection Committee of the IEEE ComSoc. Dr. Shen served as the Editor-in-Chief of the IEEE Internet of Things Journal, IEEE Network, and Peer-to-Peer Networking and Applications.
\end{IEEEbiography}


\vspace{12pt}
\end{CJK}
\end{document}